\documentclass[usenatbib]{mnras}

\usepackage{newtxtext,newtxmath}

\usepackage[T1]{fontenc}
\usepackage{ae,aecompl}

\usepackage{mathtools}
\usepackage{siunitx}
\usepackage{multirow,tabularx}
\usepackage{amsmath}
\usepackage{graphicx}
\usepackage{comment}
\usepackage[normalem]{ulem}
\usepackage{BibDef}
\usepackage[dvipsnames]{xcolor}
\usepackage{hyperref}
\usepackage{url}
\usepackage{multirow}
\usepackage{pifont}
\usepackage{makecell}
\hypersetup{
	colorlinks=true,        
	linkcolor=blue,         
	citecolor=blue,         
}

\graphicspath{{figures/}}

\newcommand{\msun}{~\rm M_{\large \odot}}

\newcommand{\code}{\textsc{mordor}}

\defcitealias{Genel_et_al_2015}{G15}
\defcitealias{Du_et_al_2021}{D21}

\title[Mordor]{Morphological decomposition of TNG50 galaxies: methodology and catalogue}

\author[Tommaso Zana et al.]{
Tommaso Zana,$^{1}$\thanks{E-mail: tommaso.zana@sns.it}
Alessandro Lupi,$^{2}$
Matteo Bonetti,$^{2,3}$
Massimo Dotti,$^{2,3}$
\newauthor
Yetli Rosas-Guevara,$^{4}$
David Izquierdo-Villalba,$^{2,3}$
Silvia Bonoli,$^{4,5}$
Lars Hernquist,$^{6}$
\newauthor
Dylan Nelson$^{7}$
\\
$^{1}$Scuola Normale Superiore, Piazza dei Cavalieri 7, I-56126 Pisa, Italy\\
$^{2}$Dipartimento di Fisica G. Occhialini, Università di Milano-Bicocca, Piazza della Scienza 3, IT-20126 Milano, Italy\\
$^{3}$INFN, Sezione di Milano-Bicocca, Piazza della Scienza 3, IT-20126 Milano, Italy\\
$^{4}$ Donostia International Physics Centre (DIPC), Paseo Manuel de Lardizabal 4, 20018 Donostia-San Sebastian, Spain\\
$^{5}$IKERBASQUE, Basque Foundation for Science, E-48013, Bilbao, Spain\\
$^{6}$Harvard-Smithsonian Center for Astrophysics, 60 Garden Street, Cambridge, MA, 02138, USA\\
$^{7}$Institut f\"ur Theoretische Astrophysik, Zentrum f\"r Astronomie Universit\"at Heidelberg Albert-Ueberle-Str. 2, 69120, Heidelberg, Germany\\
}

\date{Accepted XXX. Received YYY; in original form ZZZ}

\pubyear{2021}

\begin{document}
\label{firstpage}
\pagerange{\pageref{firstpage}--\pageref{lastpage}}
\maketitle

\begin{abstract}
    We present \code{} (MORphological DecOmposeR, a new algorithm for structural decomposition of simulated galaxies based on stellar kinematics. The code measures the properties of up to five structural components (a thin/cold and a thick/warm disc, a classical and a secular bulge, and a spherical stellar halo), and determines the properties of a stellar bar (if present). A comparison with other algorithms presented in the literature yields overall good agreement, with \code{} displaying a higher flexibility in correctly decomposing systems and identifying bars in crowded environments (e.g. with ongoing fly-bys, often observable in cosmological simulations). We use \code{} to analyse galaxies in the TNG50 simulation and find the following: ($i$) the thick disc component undergoes the strongest evolution in the binding energy-circularity plane, as expected when disc galaxies decrease their turbulent-rotational support with cosmic time; ($ii$) smaller galaxies (with stellar mass, $10^{9} \lesssim M_{*}/\msun \leq 5 \times 10^{9}$) undergo a major growth in their disc components after $z\sim 1$, whereas ($iii$) the most massive galaxies ($5 \times 10^{10} < M_{*}/\msun \leq 5\times10^{11}$) evolve toward more spheroidal dominated objects down to $z=0$ due to frequent gravitational interactions with satellites; ($iv$) the fraction of barred galaxies grows rapidly at high redshift and stabilizes below $z\sim 2$, except for the most massive galaxies that show a decrease in the bar occupation fraction at low redshift; ($v$) galaxies with $M_{*} \sim 10^{11}~\msun$ exhibit the highest relative occurrence of bars at $z=0$, in agreement with observational studies. We publicly release \code{} and the morphological catalogue of TNG50 galaxies.   
 \end{abstract}




\section{Introduction}
\label{sec:intro}

In addition to pure discs and systems supported by isotropic velocity distributions, a number of different galactic constituents have been identified. Already in the Hubble classification disc galaxies are classified depending on the presence (or absence) of stellar bars, while more recently the presence of secular bulges such as rotationally supported pseudo-bulges, boxy-peanut bulges and X-shaped bulges is further extending the complexity of the subjects.
These structures can also contribute to the understanding of galaxy evolutionary pathways
\citep[e.g.][]{Athanassoula_2003, Kormendy_Kennicutt_2004, Sellwood_2014, George_et_al_2020, Geron_et_al_2021}.

In the era of integral field spectroscopy the decomposition into different morphological components should, in principle, make use of all available information, including surface brightness morphology and kinematics. 
Galaxy modeling considering both photometric and spectroscopic data can be obtained using, e.g., the Schwarzschild method (\citealt[][]{Schwarzschild_et_al_1979}, see \citealt{Jethwa_et_al_2020, Vasiliev_Valluri_2020} for recent implementations of the method).
However, only individual cases have been analyzed in full detail \citep[see, for example,][]{Zhu_et_al_2018b}. 
Most of the available morphological decompositions are still based on photometry only \citep[e.g.][]{Simien_1989, Reese_et_al_2007, Bottrell_et_al_2019, Lingard_et_al_2020, Mendez-Abreu_et_al_2021, Rios-Lopez_et_al_2021}.
Moreover, current limitations of modern observations contribute to reducing our understanding of the evolution of the galaxies being studied.
As an example, the dependence of the bar occurrence frequency as a function of galaxy stellar mass and redshift can be strongly biased by the observational sample used \citep[see, e.g., the discussion in][and references therein]{Erwin_2018}.

The observational effort is supported by a parallel numerical endeavour. 
State of the art cosmological simulations of galaxy formation and evolution \citep[see][]{Vogelsberger_et_al_2020}, such as Illustris \citep{Vogelsberger_et_al_2014a}, Horizon-AGN \citep{Dubois_et_al_2014}, Eagle \citep{Schaye_et_al_2015}, TNG300/100 \citep[][]{Pillepich_et_al_2018a}, TNG50 \citep{Nelson_et_al_2019a, Pillepich_et_al_2019}, or NewHorizon \citep{Dubois_et_al_2021}, are indeed able to produce a morphological mix of galaxies that is in good agreement with well-established observational results. Noticeably, the simulated galaxies are affected by completely different selection biases with respect to their observed counterparts: the 3-D mass distribution of stars is known at any given redshift and is not affected by issues related to stellar ages and obscuration, nor by the lack of angular resolution affecting high redshift galaxies in the observational regime.  Simulations are, however, intrinsically limited by their spatial (and mass) resolution (in addition to the underlying physics assumed in the model), and may fail in forming and correctly evolving structures that are not properly resolved.

Although most of the large cosmological simulations discussed above have a typical spatial resolution of $\sim1$~kpc (with the exception of TNG50 and NewHorizon, reaching a fraction of a kpc), often insufficient for properly resolving bulges, bars and other nuclear structures, some recent studies attempted nevertheless to quantify the structural properties of galaxies in these simulations.
For instance, \citet{Dubois_et_al_2016} and \citet{Deeley_et_al_2021} focussed on galaxy morphology: the former by studying how feedback from massive black holes in the Horizon-AGN run affects the morphological appearance of galaxies, the latter by analysing the formation process of S0 galaxies in the TNG100 simulations.
\citet[][]{Pulsoni_et_al_2020} investigated, the kinematic and photometric properties of simulated early-time galaxies from TNG50 and TNG100, whereas \citet{Jagvaral_et_al_2022} proposed two strategies to decompose discs and spheroids in TNG100 galaxies.
\citet[][in EAGLE]{Algorry_et_al_2017}, \citet[][in Illustris]{Peschken_Lokas_2019}, \citet[][in TNG100]{Rosas-Guevara_et_al_2020}, \citet[][in Illustris and TNG100]{Zhou_et_al_2020}, \citet[][in NewHorizon]{Reddish_et_al_2022}, and \citet[][in TNG50]{Rosas-Guevara_et_al_2022} analysed the properties of bars, whereas \citet[][]{Gargiulo_et_al_2022} studied how the environment and the presence of bars can affect the S\'ersic profile of bulges in TNG50 galaxies.
Aside from the work of \citet{Reddish_et_al_2022}, these studies only considered disc galaxies with stellar masses of at least $\sim 10^{10}~\msun$, mostly because of the resolution limitations of the above-mentioned runs.

Any detailed analysis and/or comparison with observational data requires a proper morphological decomposition of simulated galaxies.
Recently, some studies have addressed this need.
In particular, \citet[][hereafter \citetalias{Genel_et_al_2015}]{Genel_et_al_2015}, \citet{Du_et_al_2019}, \citet{Du_et_al_2020}, and \citep[][hereafter \citetalias{Du_et_al_2021}]{Du_et_al_2021} applied two different methods to decompose the galaxies resulting from the IllustriTNG simulations.
\citetalias{Genel_et_al_2015} based their analysis on the distribution of stellar circularities alone, in order to disentangle rotation-supported sub-structures from spheroidal ones.
This fast procedure has been applied to all the galaxies in the entire IllustrisTNG project, giving a basic estimate of the disc and bulge masses.
On the other hand, \citet{Du_et_al_2019}, \citet{Du_et_al_2020}, and \citetalias{Du_et_al_2021} developed an automated scheme based on the GMM algorithm (\emph{Gaussian Mixture models}) by \citet{Obreja_et_al_2018} \citep[see also][]{Domenech-Moral_et_al_2012}, and applied it to the most massive rotation-dominated TNG100 galaxies at $z=0$ (note also that \citealt{Du_et_al_2019} focussed only on unbarred galaxies).
This method still relies on the computation of the kinematic properties of the stellar particles (i.e. the binding energy and the angular momentum components). Particles are assigned to a varying number of morphological components of Gaussian shape via an unsupervised machine learning algorithm.

In this work, we present a detailed morphological decomposition of all the galaxies within the highest resolution run of the IllustrisTNG suite, TNG50, using an automated analysis that does not require any visual inspection.
%
The algorithm is completed by a routine that identifies and characterises bars within galaxies, whenever they are present.
We remark that, the procedure employed to identify the different galactic constituents relies on stellar kinematics, whereas only the module to assess the bar presence is based on a pure morphological analysis of the stellar surface density.
However, although both ``cold'' and ``warm' are usually employed to describe the related kinematic disc components (see, e.g., \citet{Du_et_al_2020}), we will adopt hereon their morphological counterparts, i.e. ``thin'' and ``thick'', to refer to the observational appearance of the structures \citep[see][for further details]{Zhu_et_al_2018a}.

Our decomposition inherits the results of (i) \citet{Abadi_et_al_2003}, who for the first time used the circularity parameter $j_{z}/j_{\rm circ}$, i.e. the ratio between the vertical component of the angular momentum $j_{z}$ and the maximum angular momentum $j_{\rm circ}$ (that of a circular orbit) with the same total specific energy, in order to probe which systems are dispersion-dominated and which are mainly rotationally-supported \citep[see an example of application in][]{Tissera_et_al_2012}; and of (ii) \citet{Domenech-Moral_et_al_2012}, who introduced an important dependence on the binding energy of the stars.
The mass elements are then associated with a given component depending on circularity and energy cuts, as done in \citet{Du_et_al_2019} and \citet{Du_et_al_2020} for the TNG100 simulation.

A detailed description of the method is presented in Section~\ref{sec:method}, whereas the results, along with a comparison with previous findings, are discussed in Section~\ref{sec:results}.
We finally summarise our findings in Section~\ref{sec:conclusions}.


\section{Methodology}
\label{sec:method}

Here we describe our kinematic decomposition (MORphological DecOmposeR, \code{}) method which we apply to all the galaxies of the TNG50 simulation that have been identified as ``cosmological objects'' by \textsc{subfind} and have a number of stellar particles larger than $10^4$, corresponding to a minimum stellar mass\footnote{In this work, we define the total stellar mass of a galaxy $M_{*}$ as the sum of the mass of all the stellar particle assigned by \textsc{subfind} to the galaxy.} $M_{*} \sim 10^{9}\msun$. Although this choice is arbitrary, we expect smaller galaxies to be mostly irregulars (except for some dwarf ellipticals), and more importantly we do not expect their stellar dynamics to be well resolved in TNG50 -- because of the mass resolution limit -- making the identification of rotational structures potentially inaccurate. We note however that our decomposition tool, which is publicly available at \url{https://github.com/thanatom/mordor}, can work at all mass resolutions, and can therefore be applied to any cosmological simulation output obtained with any available code.

\subsection{TNG Simulations}

The IllustrisTNG suite \citep{Nelson_et_al_2018, Naiman_et_al_2018, Pillepich_et_al_2018b, Marinacci_et_al_2018, Springel_et_al_2018} includes three main cosmological, gravo-magneto-hydrodynamical simulations of galaxy formation and evolution in cubes of side $300$, $100$, and $50$~cMpc.
Each run -- with increasing mass and spatial resolutions, as the simulation box is reduced -- has been performed with the moving-mesh code \textsc{arepo} \citep{Springel_2010} and assumes a flat {$\Lambda$}CDM model with the parameters from \citet{Planck_2016}, i.e. ${\Omega_{\rm M,0}= 0.3089}$, ${\Omega_{\rm \Lambda,0}= 0.6911}$, ${\Omega_{\rm B,0}= 0.0486}$, ${H_0 = 67.74~\rm{km~s}^{-1}~{\rm Mpc}^{-1}}$. In addition to gravity, the simulations capture the physics pertaining to several baryonic processes through sub-grid modelization based on the precursor project Illustris \citep[][]{Vogelsberger_et_al_2014a, Vogelsberger_et_al_2014b, Genel_et_al_2014}, although numerous improvements have been made to stellar and black hole physics prescription as detailed in \citet{Weinberger_et_al_2017} and \citet{Pillepich_et_al_2018a}.

In this work we focus on TNG50 \citep{Nelson_et_al_2019a, Pillepich_et_al_2019} which is the highest resolution run of the suite, with a box of $50$~cMpc side, a spatial resolution for the stellar elements of $576$~ckpc down to $z=0.5$ and $288$~kpc at later times, and a mass resolution of $8.5\times10^{4}\msun$.

\subsection{Kinematic components: energy and circularity}

The kinematic decomposition is based on two main quantities associated with the particles used to sample a galactic system: the total specific energy ($E$) and the circularity ($\eta$). Whereas the total energy is easily determined from the particle velocity $\mathbf{v}$ and the gravitational potential $\psi$ as $E=1/2|\mathbf{v}|^2 + \psi$, the circularity $\eta$ can be determined in two different ways.
The easiest and computationally cheapest way to define the circularity is $\eta \equiv j_z/j_{\rm circ}(R)$, where $j_z$ is the angular momentum along the $z$ axis, and $j_{\rm circ}(R) \equiv R \ v_{\rm circ}(R)$ is the angular momentum of a circular orbit with radius $R$ in the galactic mid-plane; i.e. the maximum achievable angular momentum. Here, $R$ is the cylindrical radius of the particle, and $v_{\rm circ}=\sqrt{{\rm G}M(<\!R)/R}$ the corresponding circular velocity, estimated under the assumption of a spherically symmetric distribution \citep[see, for example,][]{Marinacci_et_al_2014}.
Despite its computational efficiency, this approach can lead to serious inconsistencies, depending on galaxy structure and the relative importance of the axisymmetric components over spherically-symmetric ones.

%
Therefore, in this work we employ a more accurate approach, using $\eta \equiv j_z/j_{\rm circ}(E)$, where $E$ the orbital energy and $j_{\rm circ}(E)$ is computed under the assumption of axial symmetry in the equatorial plane of the potential. 
As a consequence, our decomposition method also requires the determination of the gravitational potential for the evaluation of the circularity for all the selected stellar particles.

\subsection{Gravitational potential evaluation}

The computation of the gravitational potential represents a crucial aspect of \code{}, since it is by far the most time-consuming and resource-demanding section of our algorithm.
In principle, when the gravitational potential at the particle locations is stored in the galaxy snapshots (as it is the case for TNG50 20 ``main snapshots''), the kinematic decomposition can be directly performed at a negligible computational cost.
However, this information is not available for the remaining 80 snapshots. 
In this work, in particular, we rely on the public version of \textsc{arepo} to estimate the potential for each selected galaxy.
The reasons behind the choice of recomputing the potential for every object using \textsc{arepo} are manifold: 
(i) the code is the same one adopted for the simulation and thus maximizes consistency in the calculations, especially in the implementation of the gravitational softening (which is not straightforward in other methods); 
(ii) the efficient parallelization provides the best performance (among the procedures we tested), especially because of the large number of galaxies decomposed, each one with a number of particles ranging from $\sim10^{4}$, up to $\sim10^{9}$;
(iii) although the motion of each particle depends on the total gravitational potential of the simulation, thus on the position and mass of every particle in the cosmological volume (including satellites and filaments), the morphology of the galaxy is mainly defined by its own self-gravity; this allows us to compute the potential as if the galaxy was in isolation; i.e. accounting only for the potential generated by particles actually belonging to the galaxy. Moreover, the assumed isolation naturally removes possible disturbed energy distributions resulting from the cosmological interplay among galaxies.
Nevertheless, to make the tool more general, in the publicly released version we provide alternative and self-contained approaches to determine the galaxy potential, which do not require the use of \textsc{arepo} (see Appendix~\ref{sec:potential_details}).

\subsection{Decomposition algorithm}
\label{subsec:decomp_alg}

By splitting the stellar particles in different regions in the energy-circularity phase space, we identify five different components: a thin disc, a thick disc, a pseudo-bulge, a central spheroid (dubbed ``bulge''), and a less bound stellar spheroidal halo. Although this five-component approach is similar to that in \citetalias{Du_et_al_2021}\footnote{We note that, in \citetalias{Du_et_al_2021} the five components are defined as cold disc, warm disc, disky bulge, bulge, and halo, respectively.}, our method is not based on a Gaussian fitting machine learning algorithm, but on the procedure detailed in the following:
\begin{enumerate}
    \item for each galaxy identified by \textsc{subfind} that fulfills our minimum particle number criterion, we extract the entire particle list (gas, dark matter, stars, and black holes) from TNG50 snapshots, and store all the information in an HDF5 file with the same format of the \textsc{arepo} outputs;
    \item we correct for coordinate periodicity (if necessary) and compute the gravitational potential using \textsc{arepo} (see Appendix~\ref{sec:potential_details} for details).
    \item we load the produced output with \textsc{pynbody} \citep{Pontzen_et_al_2013}, and recentre the galaxy at the origin of the coordinate system via a shrinking sphere algorithm starting at the potential minimum of the entire system.
    In some cases, the stellar component happens to be significantly decoupled from the denser region of the DM halo, resulting in a wrong centring.
    In order to overcome this issue, we move to the reference frame of the centre of mass just determined, and repeat the centre calculation using the stellar component only. In case the newly found centre lies outside $\textrm{max}[2.8h(z), 0.5r_{\rm hm}]$ -- with $r_{\rm hm}$ the stellar half-mass radius and $h(z)$, the gravitational softening at the redshift $z$ -- the galaxy is shifted to this new reference frame.
    After centring, we compute the angular momentum of the galaxy from the stellar particles within 3 times the stellar half-mass radius\footnote{Whereas taking a fraction of the virial radius of the halo would have been much easier, during the analysis we noticed that some galaxies in the TNG50 catalogue exhibit a subdominant DM component, likely because they are satellites of a much larger halo and the dark matter particles failed to be associated to the satellite and were linked to the central galaxy of the halo.
    For those galaxies, neither the virial radius of the halo from the catalogue, nor the maximum distance of the dark matter particles give a proper estimate of the actual size of the system.
    The choice of using, instead, a multiple of the stellar half-mass radius is dictated by the fact that particles at very large distances from the centre would strongly affect the angular momentum determination, likely producing inaccurate alignment in the case of kinematically decoupled streams/tails of stars resulting from mergers or stripping events.} $r_{\rm hm}$ and align it with the $z$-axis;
    \item We estimate the circular angular momentum and total energy profiles in the galactic plane -- defined by the radius $R=\sqrt{x^2+y^2}$ and null $z$-coordinate - that are needed to sample $j_{\rm circ}(E)$ for the stellar particles, as $j_{\rm circ}(R,0) \equiv Rv_{\rm circ}(R,0)$ and $E_{\rm circ}=1/2v_{\rm circ}^2+\psi$.
    
    The calculations are performed via sampling 100 logarithmically-spaced radii $R$, starting from the galaxy centre up to the distance of the farthest particle in the galaxy. For each bin, we compute the potential and the gravitational acceleration via direct summation at the four $(x,y,z)$ positions $(R,0,0), (0,R,0), (-R,0,0), (0,-R,0)$, and average them to minimize the impact of the exact particle distribution.
    Finally, we use the two profiles to parameterise $j_{\rm circ}(E_{\rm circ})$, which is employed to interpolate over the stellar particle energy distribution \citep[this follows the approach already available in \textsc{pynbody} for morphological decomposition;][]{Pontzen_et_al_2013}.
    
    \begin{figure*}
    \centering
    \includegraphics[width=0.83\textwidth]{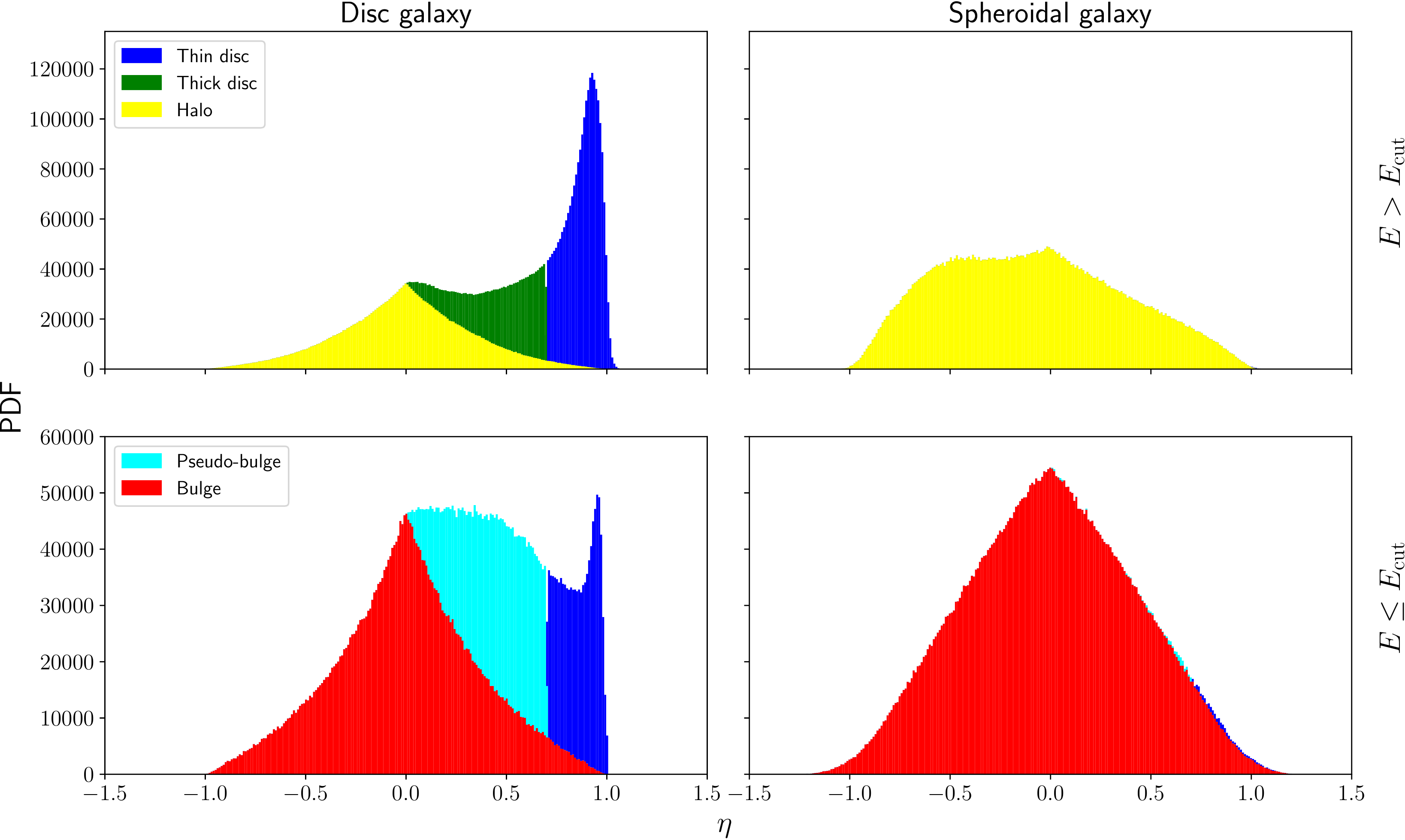}\\
    \caption{Mass weighted PDFs of circularity for DG ($D/T=0.59$) with $M_{*} = 3.8\times10^{11}$ ({\it left column}) and SG ($D/T=0.07$) with $M_{*} = 1.4\times10^{11}$ ({\it right column}) at $z=0$.
    Particles in the {\it top row} have $E>E_{\rm cut}$, whereas {\it bottom row} shows particles with $E \leq E_{\rm cut}$. The different components are selected by requiring a minimum $\eta=0.7$ for the thin disc (blue), a distribution centred on $\eta=0$ for the bulge (red) and halo (yellow) and by identifying the mildly rotating component as thick disc (green) and pseudo-bulge (cyan) respectively.}
    \label{fig:Spiral_Elliptical_dist}
    \end{figure*}

    \item At this point, in order to remove possible kinematically decoupled or unbound structures that would spuriously affect our decomposition, we exclude all stellar particles having $E\geq0$, $\lvert\eta\rvert\geq1.5$, or $\lvert j_{p}/j_{\rm circ}\rvert\geq1.5$ \citep[][]{Du_et_al_2019} from the analysis, where $j_{p}$ is the component of the angular momentum in the $x-y$ plane, i.e. $j_{p}=\sqrt{j_{\rm x}^2 + j_{\rm y}^2}$.
    We note that this choice removes, at most, 1-2\% of the particles, hence it does not significantly affect the results.
    \item Next, we bin the particle energy distribution in 25 bins ranging from $E_{\rm min}$ to the $90$th percentile of the distribution\footnote{As a first attempt, we avoid the whole energy distribution since, in many cases, close-to-being unbound satellites or tidal streams which can deceive the algorithm can be found around galaxies.}, and look for any relative minimum in the distribution, requiring that,
    for each bin $i$, (A) $N_{i-1}>N_i$ and $N_{i+1} \geq N_i$ or $N_{i-1} \geq N_i$ and $N_{i+1}>N_i$, with $N_i$ the number of particles in the $i$-th bin, (B) $N_{i-2}>N_i$ and $N_{i+2}>N_i$, (C) $\sum_{i}^{x>i} {N_{x}} > N_{\rm min}$, where $N_{\rm min}=\max(1000,0.01N_\star)$ and $N_\star$ is the total number of stellar particles considered.
    This last requirement ensures that the least-bound components have a sufficient number of particles to reliably perform the subsequent analysis.
    If no minima are found, the search is repeated extending the analysis to the entire energy distribution.
    If, even in this case, a clear minimum cannot be found, we assume that only three components exist in the galaxy and, for simplicity, we identify them as the most bound ones, i.e. bulge, pseudo-bulge, and thin disc.
    If, instead, one or more minima are found, we iteratively repeat the search, doubling the number of bins at each iteration, and requiring that each new minimum lies at a distance $d<3\Delta E_{\rm bin}$ from one of the previously found minima, with $\Delta E_{\rm bin}$ the bin size at the current iteration step.
    The iteration is stopped either when the maximum number of bin is reached\footnote{We select the maximum bin number as the integer part of $\sqrt{N_\star}/2$, but only in the range between 80 and 400, where $N_\star$ is the total number of bound stellar particles. We found this choice to minimize the numerical noise resulting from poorly filled bins.}, or when a single minimum is found.
    In case the iteration is interrupted before reaching the maximum number of bins, the position of the single minimum found is refined further.
    At the end of this procedure, we split the particle distribution according to the lowest minimum found, dubbed $E_{\rm cut}$, into more bound ($E \leq E_{\rm cut}$) and less bound components, whereas, if no minima have survived the selection, three single components are again assumed.
    \item We then proceed analysing the most bound components, binning their particles by circularity.
    All stellar particles with $\eta<0$ are directly assigned to the bulge, and, since we expect the spheroidal components to be symmetric about $\eta=0$, we select an equivalent distribution from the positive circularity part of the distribution. We also flag the particles to be assigned to the bulge via a Monte-Carlo sampling. 
    Among the remaining (not assigned to the bulge) particles, we then identify those with $\eta>0.7$ as belonging to the thin disc component, and the remainder are associated to a pseudo-bulge component. The same procedure is applied to the less bound particles (if present), assigning them to the stellar halo (in the same way as the bulge), to the thin disc, or to a thick disc.
    Fig.~\ref{fig:Spiral_Elliptical_dist} exemplifies how particles are assigned to different components according to their circularity in our procedure. In particular, we show the results for two quite massive galaxies at $z=0$: a disc galaxy (ID: 360923, hereafter dubbed DG; left panels) and a disc-to-total ratio $D/T\equiv M_{\rm disc}/M_{*} = 0.59$, where $M_{\rm disc}$ is defined as the cumulative mass of all the rotating components (i.e. the thin disc, the thick disc and the pseudo-bulge, see \S~\ref{subsec:redshift_evolution}), and a spheroidal galaxy with $D/T=0.07$ (ID: 253863, hereafter dubbed SG; right panels). The ratio of the thin disc component alone is even more diverse between the two systems, being $M_{\rm thin}/M_{*}=0.36$ and $0.02$ for DG and SG, respectively.
    In the probability distribution functions (PDFs), the disc component is clearly visible in the left panels at high circularities, whereas only a trace is present in SG, on the right.
    The dispersion-supported components (bulge and halo) are perfectly centred on $\eta=0$, thus demonstrating the accuracy of our identification algorithm.
    We also note that the less bound component in SG is not perfectly symmetric, showing a slightly larger amount of mass with $\eta<0$.
    This could be due to a small number of counter-rotating particles which we assign to the halo as well.
    \item Finally, the total mass and the average energy and circularity of each component are estimated by summing up the properties of the identified stellar particles.
    The results are reported in a publicly available catalogue.\footnote{\url{www.tng-project.org/zana22}}
    An important caveat in this analysis is that, if bar structures are present, stellar particles belonging to the bar could be either associated to the pseudo-bulge or to the spheroidal component, because of their relatively low angular momentum relative to the disc. This is why, at the end of the kinematic decomposition, we perform a separate bar identification analysis on all the decomposed galaxies, that is also reported in the catalogue.  
\end{enumerate}

\begin{figure*}
    \centering
    \includegraphics[width=0.83\textwidth]{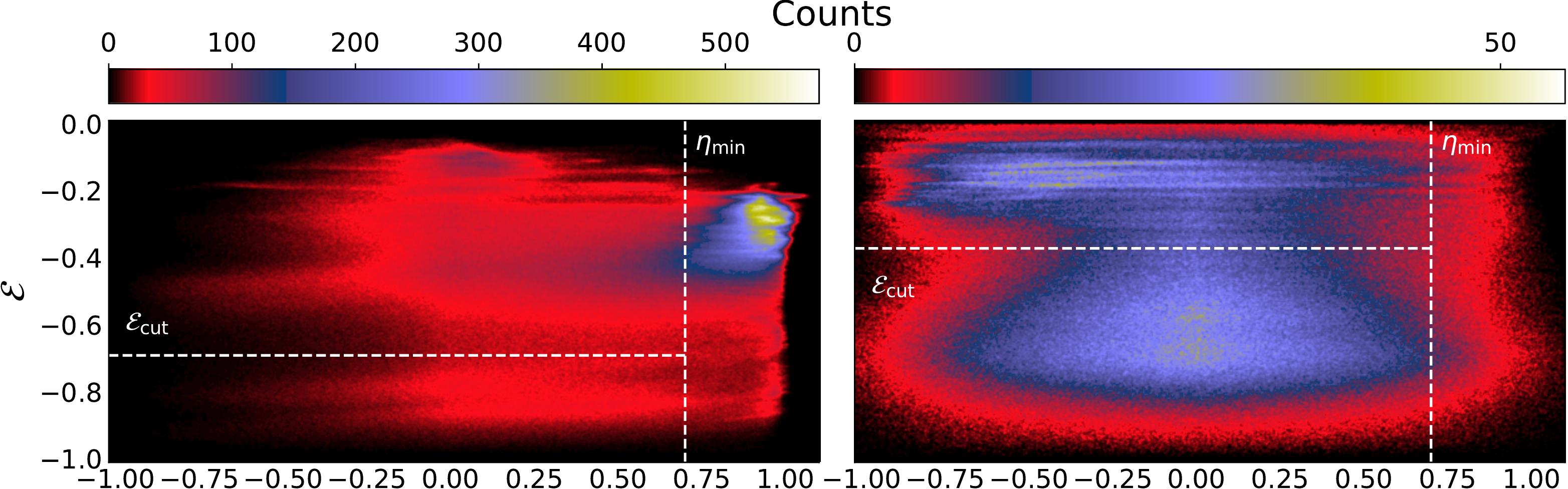}\\
    \vspace*{0.5cm}
    \includegraphics[width=0.83\textwidth]{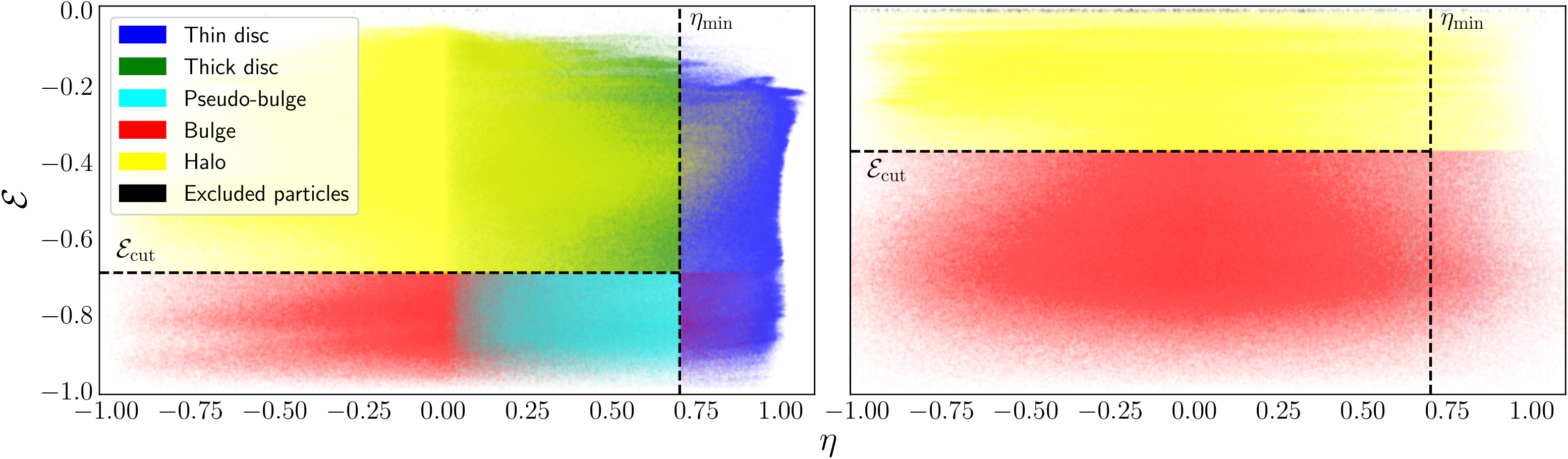}
    \caption{{\it Top row}: particle distribution in the phase space ($\eta-\mathcal{E}$) for the same galaxies shown in Fig.~\ref{fig:Spiral_Elliptical_dist}.
    Dashed white lines mark the thresholds used to separate the different components: the vertical line is set to $\eta=0.7$ a priori, whereas $\mathcal{E}_{\rm cut} \equiv E_{\rm cut}/|E|_{\rm max}\simeq 0.7$ for DG and $\mathcal{E}_{\rm cut}\simeq 0.4$ for SG are calculated through the algorithm discussed in \S~\ref{subsec:decomp_alg}.
    {\it Bottom row}: the different morphological components, i.e. thin disc (blue), thick disc (green), pseudo-bulge (cyan), bulge (red), and stellar halo (yellow) are identified for DG and SG.
    Black points mark those particles which have been rejected in our procedure because they either are unbound or exhibit a too high perpendicular/parallel angular momentum.}
    \label{fig:Spiral_Elliptical}
\end{figure*}

Fig.~\ref{fig:Spiral_Elliptical} shows the final result of our procedure.\footnote{In the interpolation procedure we adopt to evaluate $j_{\rm circ}$, the finite number of bins could exclude some particles from the computation, having $E>E_{\rm circ, max}$, where $E_{\rm circ,max}$ is the circular energy in the last bin of the distribution. In order to avoid these outliers (which are rarely more than a few percent of the total) we force these particles to be part of the spheroidal components, either placing them in the last bin of the energy distribution, or by setting their circularity to zero. In Fig.~\ref{fig:Spiral_Elliptical} we follow the first method.} 
In the top row, the differences are clear: whereas most of the stellar particles in DG have $\eta \sim 1$, SG shows two massive components centred around $\eta \sim 0$, both above and below $\mathcal{E}_{\rm cut} \equiv E_{\rm cut}/|E|_{\rm max}$.
It is also clear that our method successfully identifies the natural separation in the energy distribution of the two systems, around $\mathcal{E}=0.7$ and $\mathcal{E}=0.4$ in DG and SG, respectively.

Analogously, in the bottom row, the different components are marked with the same colour code used in Fig.~\ref{fig:Spiral_Elliptical_dist}.
Despite the fixed threshold to identify the thin disc, the Fig. shows that the difference between contiguous components is not always sharp, but some density gradients are visible.

\begin{figure*}
    \centering
    \includegraphics[width=0.88\textwidth]{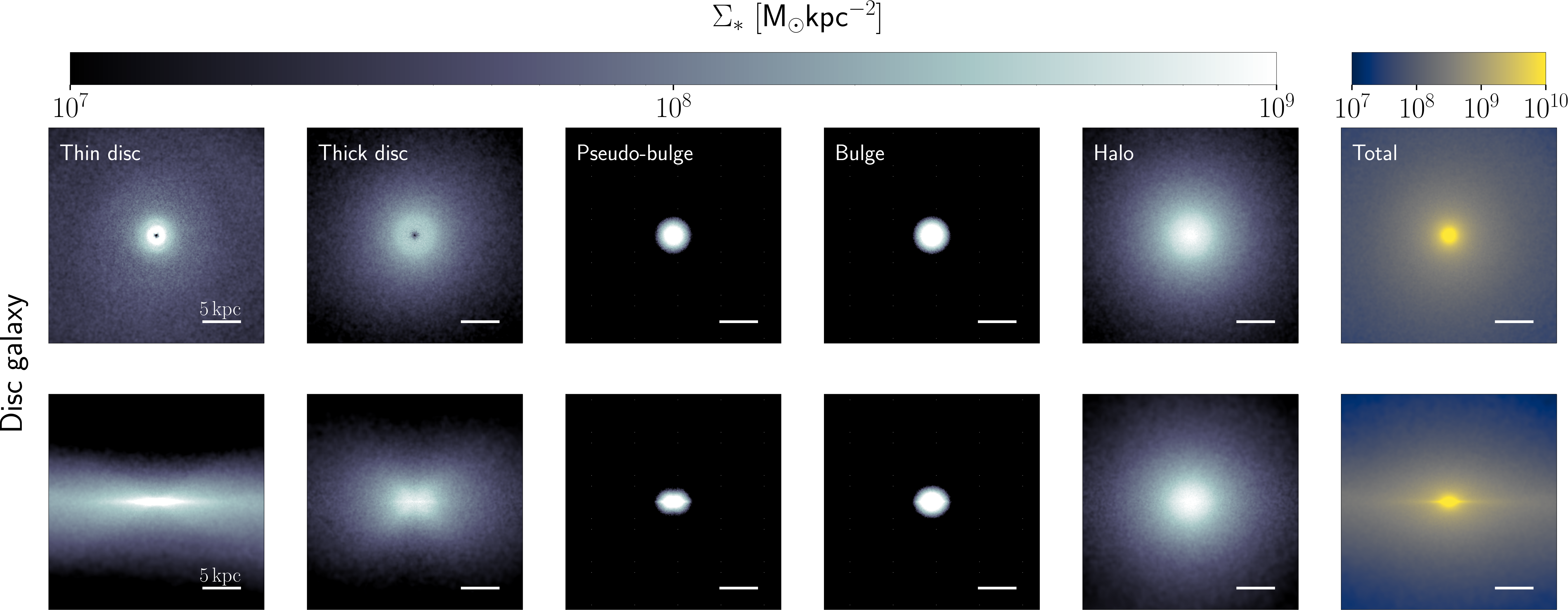}\\
    \vspace*{1.2cm}
    \includegraphics[width=0.88\textwidth]{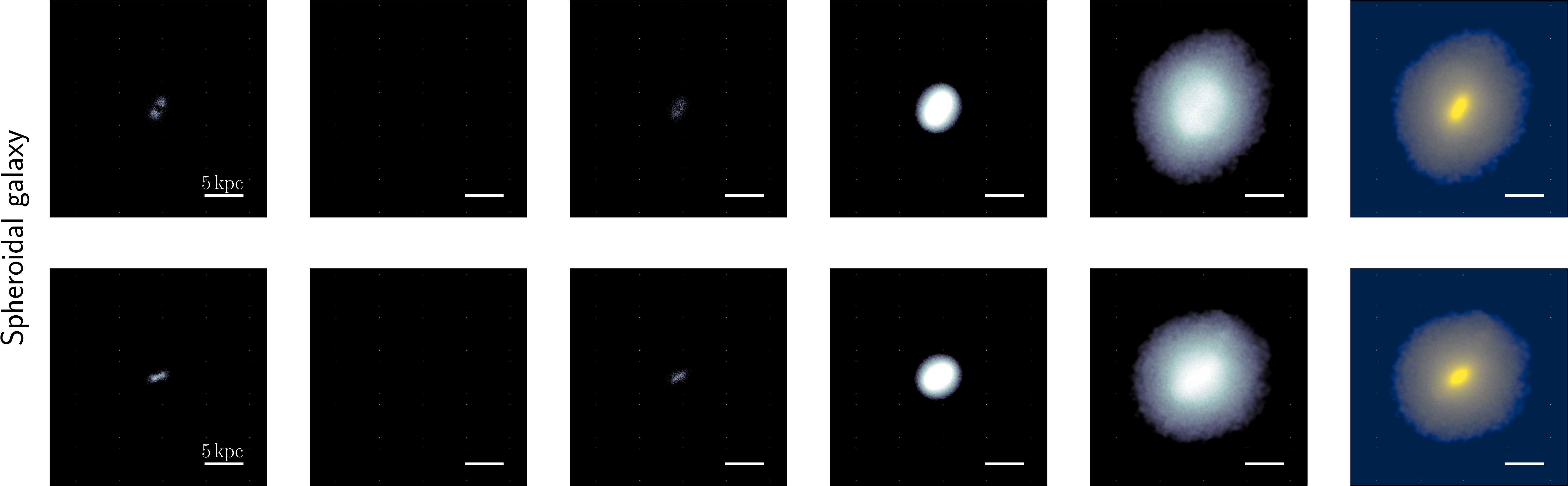}    
    \caption{Stellar surface density maps of the spiral ({\it top rows}) and spheroidal ({\it bottom rows}) galaxies decomposed in Fig.~\ref{fig:Spiral_Elliptical_dist}, seen both face-on ({\it first and third rows}) and edge-on ({\it second and forth rows}).
    From left to right we show the thin and thick discs, the pseudo-bulges, the bulges, the haloes, and the entire stellar distributions.}
    \label{fig:density_maps}
\end{figure*}

To conclude, in Fig.~\ref{fig:density_maps} we show the stellar density maps of the various components extracted from the two example galaxies, along with their total stellar distribution.
While the massive rotating components are clearly visible in the DG case, only a small trace is observed in SG.


\subsection{Bar identification}
We determine the presence (or absence) of a bar structure through a Fourier decomposition of the stellar surface density field, after the system has been rotated to align the total angular momentum with the z-axis. 
Our procedure is based on the methods developed in \citet{Zana_et_al_2018a} and \citet{Zana_et_al_2019}, although some fundamental updates are introduced here in order to improve the analysis and make it reliable even for galaxies with less prominent disc components, or with a disturbed morphology, more frequent at high redshift in cosmological simulations.
In strongly dispersion-dominated systems, the total angular momentum is small, and the alignment of the (possibly present) disc component is not accurate.
%
Although this aspect, for obvious reasons, is not relevant to the morphological decomposition (if the rotating component is small, the effect would be minimum by construction), it has a crucial importance for the Fourier decomposition.
When the alignment is not accurate, a spurious two-mode could arise in the Fourier series because of projection effects and numerical noise.
To avoid these ``fake bars'', we ensure that the rotation procedure is optimal by excluding from the analysis all galaxies that exhibit a low fraction of stellar kinetic energy in ordered rotation \citep[see][]{Sales_et_al_2010},
\begin{equation}
    \kappa_{\rm rot} =  \frac{\sum_{i} m_{i} (j_{z,i}/R_i)^{2}}{\sum_{i} m_{i} v_i^{2}},
    \label{eq:Krot}
\end{equation}
where $R_i$, $v_i$, and $m_i$ are the stellar particle cylindrical radius,  velocity, and mass, and where the summation is carried over all the stellar particles enclosed in a sphere of radius 3$r_{\rm hm}$, to be consistent with the region where we compute the galaxy angular momentum.
To be conservative and avoid the vast majority of fake bars, we set our threshold at $\kappa_{\rm rot}=0.4$.\footnote{We note that even galaxies with $0.35\lesssim \kappa_{\rm rot}<0.4$ are often correctly rotated, with small uncertainties. Nevertheless, we chose to exclude these objects since we found, by visual inspection, that these systems more frequently show elongated/oblate bulges rather than actual bars.}
Although some barred galaxies are inevitably missed with this limitation, they are not a significant fraction, since when the disc component is under-massive, bars are less likely to form \citep[see, e.g.,][]{Efstathiou_et_al_1982}.

After the galaxy is aligned, we determine the radial profile of the ratio between the second and the zeroth term in the Fourier decomposition, defined as
\begin{equation}
    A_{2}(R) \equiv \frac{\left|\sum_{i}{m_{i}e^{2i\theta_{i}}}\right|}{\sum_{i}m_{i}} ,
    \label{eq:A2}
\end{equation}
where $\theta_{i}$ is the angular position in the plane of the $i$-th particle, and the sum is performed over all the stellar particles within a shell centred at $R$.
Analogously, we define the quantity $A_{2}(<\!R)$,
for which the summation is instead performed over all the particles enclosed within $R$.

To identify the presence of a bar and estimate its strength, we rely on the (local) maximum of the $A_{2}(R)$ profile.
In principle, the analogous maximum in the $A_{2}(<\!R)$ profile could provide an alternative strength estimate, but its value is often affected by central mass concentrations in the galaxy.
Nevertheless, $A_{2}(<\!R)$ is useful in order to detect the maximum extent of the asymmetry.
In this work, we subdivide the particles into cylindrical bins, starting from the centre, up to 4 times the half mass radius of the galaxy stellar component, imposing a maximum of 200 bins and a minimum of 300 particles inside each bin, to minimize the numerical noise.

\begin{figure*}
    \centering
    \includegraphics[width=0.65\textwidth]{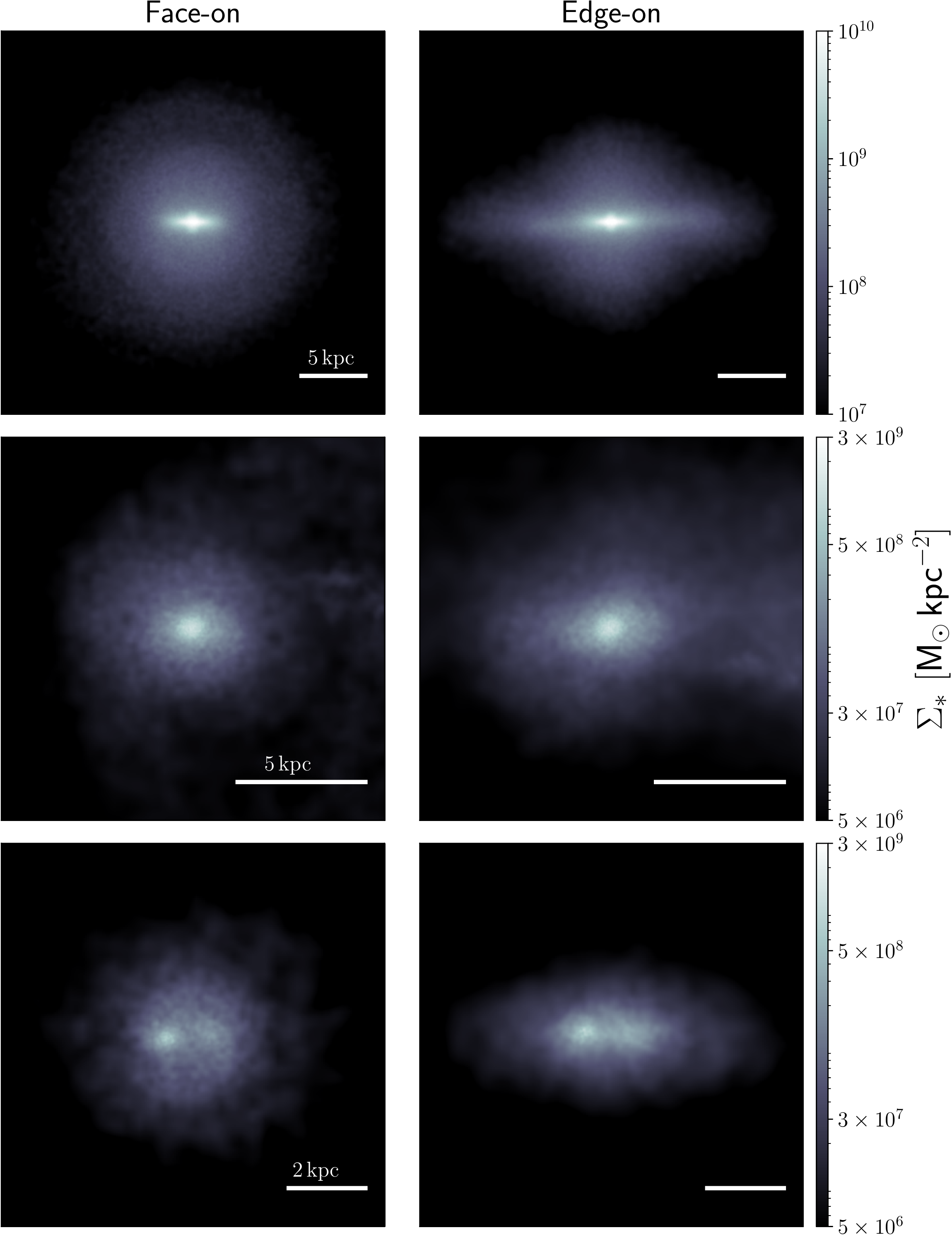}\\
    \caption{Stellar surface density maps of three candidate barred galaxies at $z=0$ face-on view ({\it left column}) and edge-on view ({\it right column}).
    The first object (ID 513105; $M_{*}=4.6\times10^{10}~\msun$; {\it top row}) is identified as a spheroidal ($M_{\rm bulge}+M_{\rm halo}=0.82M_{*}$) barred galaxy.
    The second object (ID 600743; $M_{*}=9.6\times10^{9}~\msun$; {\it middle row}) is classified as unbarred, since $\sigma_{z}/\sigma_{R}=1.3$.
    The third object (ID 738336; $M_{*}=1.3\times10^{9}~\msun$; {\it bottom row}) has a strong two-mode, but is eventually classified as an unbarred galaxy since $M_{b1}/M_{b2}=1.8.$}
    \label{fig:bar_ex}
\end{figure*}

In order to allow the code to automatically recognise bar structures among other deviations from axisymmetry, such as fly-by encounters, spiral arms or dense stellar clusters, we implement a method similar to that developed in \citet{Zana_et_al_2019}.
In particular, for each galaxy:
\begin{itemize}
    \item we first look for any peaks\footnote{Specifically, we check where the derivative of the profile with respect to the radius changes its sign.} in the $A_2(<\!R)$ profile (appropriately smoothed, with a kernel size dependent on the number of particles\footnote{If $N_{\rm bin}$ is the number of bins and $N_{\rm bins, max}=200$ is the maximum number of bins allowed, we average the $A_2(<\!R)$ profile over $5^{\frac{4}{3}(1-N_{\rm bins}/N_{\rm bins, max})}$ bins. This kernel is increased by 2 additional bins to smooth the $A_2(R)$ and $\Phi(R)$ profiles.}), whose behaviour is far less influenced by numerical fluctuations with respect to $A_2(R)$, being computed on a higher number of particles;
    \item we check the phase of the Fourier mode $\Phi(R)$, defined by
    \begin{equation}
     \Phi(R) \equiv \frac{1}{2} \arctan \left[\frac{\sum_{i}{m_{i}\sin(2\theta_{i})}}{\sum_{i}{m_{i}\cos(2\theta_{i})}} \right],
        \label{eq:phase}
    \end{equation}
    and keep as candidate bars only those systems where the phase around the peak remains almost constant across a radial interval $\Delta R > 1.4h$, where $h$ is the gravitational softening, i.e. 
    \begin{equation}
        \left| \Phi(R_{\rm peak})-\Phi(R)\right|<\Delta \Phi,
        \label{eq:phase_criterion}
    \end{equation}
    with $R_{\rm peak}$ the peak position and $\Delta \Phi=\arcsin(0.15)$;
    \item we then refine the evaluation of the candidate bar inner and outer edges, corresponding to the first and last radial bin fulfilling the phase criterion.
    We do it by recursively applying Equation~\ref{eq:phase_criterion} in the enclosed radial range $\Delta R$, each time by computing $\Delta \Phi = \sigma_{\Phi}$, where $\sigma_{\Phi}$ is the standard deviation of $\Phi(R)$ within $\Delta R$.
    This refinement is performed three times;
    \item we filter the sample of selected candidate bars, by imposing that 
    (i) the position of the peak is $R_{\rm peak}<1.5r_{\rm hm}$; 
    (ii) the position of the inner bar inner edge, $R_{\rm min}$, is $R<{\rm max}(2.8h, 0.5r_{\rm hm})$; 
    (iii) the position of the outer edge of the bar, $R_{\Phi}$ (that we adopt as the bar half-length), is $2.8h \leq R_{\Phi} \leq 4r_{\rm hm}$,\footnote{Even if the choice of upper bound seems large, galaxies with a particularly high mass concentration can have a very low $r_{\rm hm}$ and this may result in excluding some bar-like structures in exceptional under-dense discs.} and 
    (iv) the strength of the bar, defined as the maximum of the $A_{2}(R)$ profile within $\Delta R$, is $A_{2, {\rm max}}(R)\geq0.1$;\footnote{Note that this threshold is lower than what is usually considered in fully cosmological simulations ($A_{2, {\rm max}}\geq0.2$) since TNG50 has better spacial resolution and our algorithm is accurate enough to disentangle numerical noise from the presence of real structures. However, we consider the structures with $0.1\leq A_{2, {\rm max}}<0.2$ as ``proto-bars'', i.e. weak structures, likely at the beginning of the bar formation process.}
    \item finally, if more than one peak survive, we select the nearest one to the centre of the galaxy.
\end{itemize}

Differently from the kinematic decomposition, the Fourier analysis is only performed on a cylindrical slab in the centre, where the number of mass elements can be smaller than the minimum required to resolve the extent of the smoothing kernel.
In these cases, the procedure stops and the galaxy is marked as unbarred

Finally, we perform an additional filtering to the bar candidates, in order to avoid some potential misclassifications.
A bar looks exactly as an elongated central bulge and, without a detailed orbital analysis, the simple study of the density distribution of the stellar component could lead us to misinterpret a quite elongated bulge as a bar.
Moreover, secular processes continuously reduce the differences between such systems (see, e.g., the bar buckling process, or the formation of a discy bulge).
%
We also note that, both the possible failure of the galaxy alignment procedure and the misinterpretation of elongated bulges as bars are not typically considered in works which automatically identify bars in large cosmological simulations.
The reason is that the analysis is commonly restricted to systems previously classified as disc galaxies only.
%
In this work, instead -- since the definition of ``disc galaxy'' is somewhat arbitrary and, in addition, S0 galaxies are also observed to be barred (\citealt{de_Vaucouleurs_1959}; see, e.g., first row in Fig.~\ref{fig:bar_ex}) -- we perform the Fourier decomposition on the whole galaxy sample, with the only constraint on $\kappa_{\rm rot}$.

Since bars develop from the stellar disc, inheriting its vertical density profile, the vertical velocity dispersion of stars $\sigma_{z}$ is expected to be null for an infinitely thin disc, and to slowly increase as the disc thickness grows.
Furthermore, stars trapped in bar-like orbits (e.g. the $x_{1}$ family orbits), feature a larger radial velocity dispersion $\sigma_{R}$ than the almost circular orbits in a completely axi-symmetric disc.
On the other hand, perfectly dispersion-dominated systems such as classical bulges would display the same velocity dispersion in all directions because of phase-mixing, and would maintain a higher  $\sigma_{z}/\sigma_{R}$ ratio with respect to a bar.
In addition to this, a drop in the squared ratio $(\sigma_{z}/\sigma_{R})^{2}$ is usually observed as a prelude to the vertical instability processes responsible for the bar buckling \citep[][]{Raha_et_al_1991, Martinez-Valpuesta_et_al_2006, Zana_et_al_2019}. 
This suggests the use of the $\sigma_{z}/\sigma_{R}$ ratio as an indicator of the specific structure observed, in order to discriminate between the two of them.
To perform this analysis on the bar particles only, we cut a cuboid of dimensions $R_{\Phi}\times2.8h\times2.8h$, extracted from the core of the candidate bar, along the major axis.
We then mark as bars only those structures showing $\sigma_{z}/\sigma_{R}<1$, in order to exclude at least the most obvious bulges.
In the middle row of Fig.~\ref{fig:bar_ex}, we show a candidate bar rejected because of its $\sigma_{z}/\sigma_{R}$ ratio.
The analysis of the velocity dispersion ratio in the bar core allows us to probe the general behaviour of the stellar orbits, whose detailed study would be the only way to unambiguously classify bars and discriminate then from highly elliptical bulges.
Unfortunately, such a specific analysis is extremely time and resource demanding, and cannot be used to quickly classify statistically-relevant samples of galaxies.
Furthermore, orbital analysis is also prohibitive in current cosmological simulations because of their low time sampling.\footnote{An attempt to model the orbits within galaxies from cosmological simulations has been done in TNG50 by \citet{Zhu_et_al_2022}, where tracer orbits have been integrated in a fixed background potential extracted from the simulated box at $z=0$.}

Finally, even with the supplementary criterion above, some stellar over-densities could still be wrongly identified as bars in a Fourier decomposition analysis.
In particular, merger-induced transient structures -- more frequent at high redshift -- or the fly-by of small satellites orbiting very close to the nucleus can still mislead all the criteria so far discussed and be associated with a strong peak in the $A_{2}$ profile.
For this reason, we perform a further check on the density profile of the candidate bar, and require it to be essentially homogeneous; i.e. we ensure that both sides of the bar have a similar density. 
In detail, we evaluate the mass ratio $M_{b1}/M_{b2}$, where $M_{b1}$ and $M_{b2}$ are the masses of the two halves of the cuboid with length $R_{\Phi}$ (aligned with the bar) and all other dimensions equal to $2.8h$.
We then require that $M_{b1}/M_{b2}<1+\epsilon$ and $M_{b1}/M_{b2}>1/(1+\epsilon)$, with $\epsilon=0.3$.
As an example, bottom row of Fig.~\ref{fig:bar_ex} shows a galaxy where a stellar over-density is orbiting at about $1$~kpc from the centre, resulting in a spurious peak in the $A_{2}$ profile.
The galaxy is finally classified as an unbarred system via the computation of $M_{b1}/M_{b2}$.

\begin{figure}
    \centering
    \includegraphics[width=0.5\textwidth]{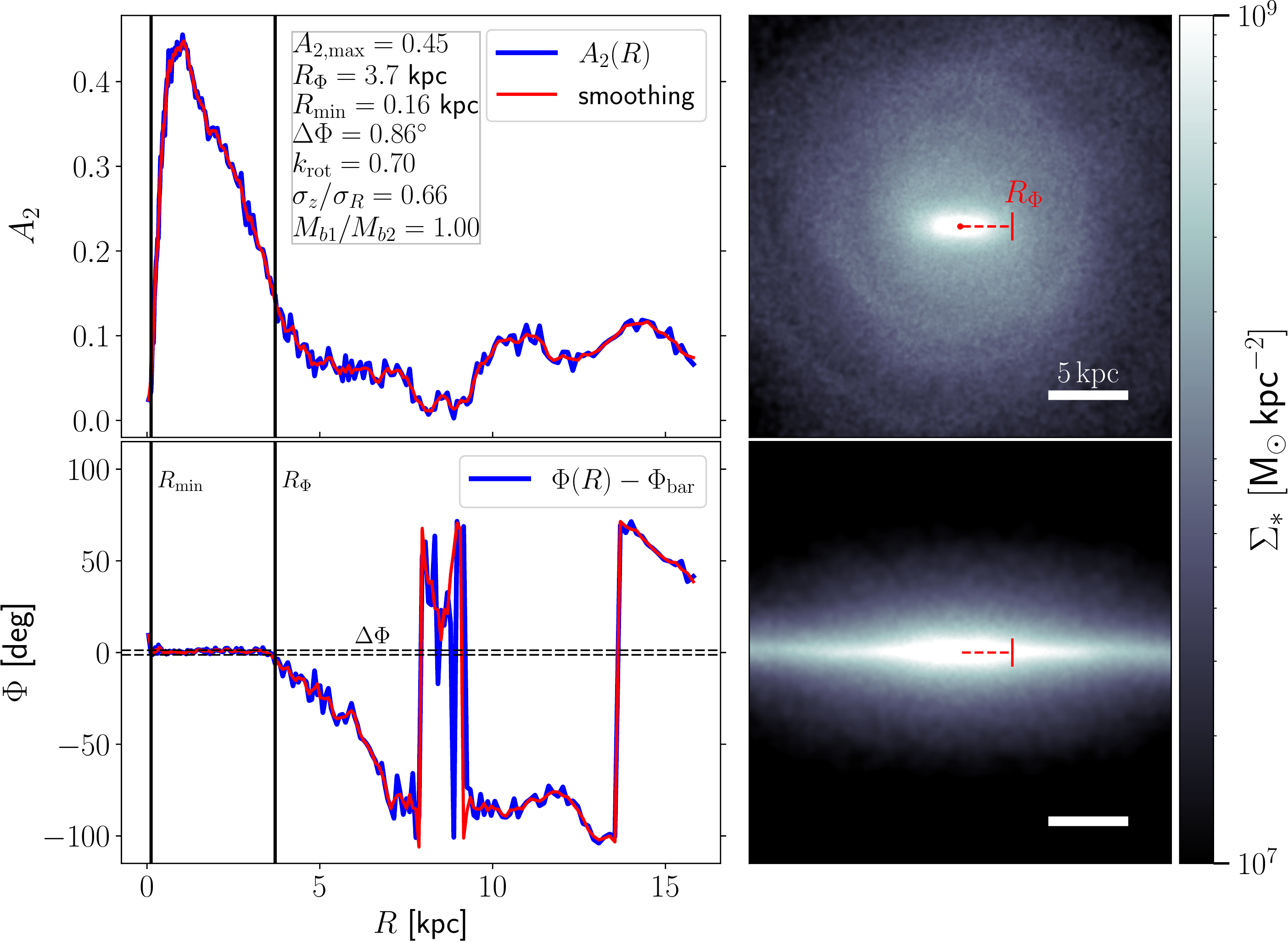}\\
    \caption{Barred galaxy 519311 (dubbed as BG) with $M_{*} = 9.2\times10^{10}~\msun$ and $M_{\rm thin}+M_{\rm thick}+M_{\rm pbulge}=0.8M_{*}$.
    {\it Left column}: profiles (blue lines) of $A_{2}(R)$, ({\it top}) and $\Phi(R)$ ({\it bottom}) along with their smoothed versions (red lines). Vertical solid black lines mark the position of $R_{\rm min}$ and of the bar extent ($R_{\Phi}$). 
    Horizontal dashed lines show the range $\Delta \Phi$ where the phase is allowed to vary within the bar.
    {\it Right column}: stellar density maps of the face-on ({\it top}) and edge-on ({\it bottom}) projections of the galaxy.
    The system is rotated in order to align the bar on the x-axis.}
    \label{fig:A2prof_ex}
\end{figure}

The results of the bar-finder algorithm are exemplified in Fig.~\ref{fig:A2prof_ex} for the barred galaxy with id 519311, hereon BG.
a
The stellar surface density maps of the face-on (top right) and edge-on (bottom right) projections of BG are shown in the right column of the Figure.
A clear bar, with an extent of almost $4$~kpc is clearly visible in the central part of the disc and in the shapes of both the $A_{2}(R)$ (top left) and the $\Phi(R)$ (bottom left) profiles.


\section{Results}
\label{sec:results}

Here, we show the results of our analysis for all the 
snapshots of the TNG50 simulation where at least a sufficiently massive galaxy has been identified (96 snapshots, from $z=9$
, to $z=0$), following the procedure outlined in the previous section.
The first 10 entries of our catalogue are listed in Table~\ref{tab:catalog_example}.
\begin{table*}
    \caption{First 10 galaxies of the catalogue at $z=0$. From top to bottom: subhalo ID, fraction of excluded particles (see text), mass fraction of thin disc, thick disc, bulge, pseudo-bulge, and halo, average circularity for all the previous morphological components, average energies of the same components, flag identifying if the subhalo host a barred galaxy or not (according to the threshold discussed in the text), bar size estimates ($R_{\Phi}$ and $R_{\rm peak}$, i.e. the position of the peak in the $A_{2}(<\!R)$ profile), bar strength estimates ($A_{2, \rm max}(R)$ and the value of $A_{2}(<\!R)$ in $R_{\rm peak}$, defined as $A_{2, \rm max}(<\!R)$), quality flags to identify real bars (fraction of rotational kinetic energy, vertical-to-radial velocity dispersion ratio, mass ratio of the two bar sides).}
    
    \centering
        \begin{tabular}{c|c|c|ccccccccccc}
        SubhaloID & & & 0 & 1 & 2 & 3 & 4 & 5 & 6 & 7 & 8 & 9 & \dots{} \\
        \hline
        \multirow{16}{*}{\rotatebox[origin=c]{90}{Morphology}} & \multirow{6}{*}{\makecell{Mass \\ fractions}} & Excluded particles & 0.000 & 0.000 & 0.000 & 0.000 & 0.000 & 0.000 & 0.000 & 0.000 & 0.00 & 0.00 & \multirow{6}{*}{\dots{}}\\
        & & Thin disc & 0.003 & 0.331 & 0.168 & 0.235 & 0.230 & 0.229 & 0.197 & 0.286 & 0.58 & 0.37\\
        & & Thick disc & 0.016 & 0.165 & 0.096 & 0.191 & 0.162 & 0.347 & 0.175 & 0.191 & 0.29 & 0.25\\
        & & Bulge & 0.827 & 0.167 & 0.109 & 0.330 & 0.378 & 0.211 & 0.391 & 0.299 & 0.03 & 0.16\\
        & & Pseudo-bulge & 0.021 & 0.082 & 0.037 & 0.108 & 0.121 & 0.069 & 0.104 & 0.092 & 0.03 & 0.10\\
        & & Halo & 0.134 & 0.255 & 0.589 & 0.135 & 0.109 & 0.144 & 0.133 & 0.133 & 0.07 & 0.12\\
        \cline{2-14}
        & \multirow{5}{*}{\makecell{Mean \\ circularities}} & $\langle \eta_{\rm thin} \rangle$ & 0.79 & 0.87 & 0.83 & 0.84 & 0.86 & 0.83 & 0.84 & 0.85 & 0.87 & 0.85 & \multirow{5}{*}{\dots{}}\\
        & & $\langle \eta_{\rm thick} \rangle$ & 0.20 & 0.50 & 0.60 & 0.51 & 0.46 & 0.46 & 0.46 & 0.49 & 0.50 & 0.49\\
        & & $\langle \eta_{\rm bulge} \rangle$ & 0.00 & 0.00 & 0.00 & 0.00 & 0.00 & 0.00 & 0.00 & 0.00 & 0.00 & 0.00\\
        & & $\langle \eta_{\rm pseudo-bulge} \rangle$ & 0.30 & 0.42 & 0.42 & 0.41 & 0.41 & 0.41 & 0.41 & 0.40& 0.44 & 0.45\\
        & & $\langle \eta_{\rm halo} \rangle$ & 0.00 & 0.00 & 0.01 & 0.00 & 0.00 & 0.00 & 0.00 & 0.00 & 0.00 & 0.00\\
        \cline{2-14}
        & \multirow{5}{*}{\makecell{Mean \\ energies}} & $\langle \mathcal{E}_{\rm thin} \rangle$ & -0.43 & -0.32 & -0.44 & -0.32 & -0.28 & -0.31 & -0.27 & -0.34 & -0.44 & -0.31 & \multirow{5}{*}{\dots{}}\\
        & & $\langle \mathcal{E}_{\rm thick} \rangle$ & -0.20 & -0.31 & -0.39 & -0.26 & -0.25 & -0.23 & -0.27 & -0.32 & -0.51 & -0.25\\
        & & $\langle \mathcal{E}_{\rm bulge} \rangle$ & -0.49 & -0.75 & -0.82 & -0.72 & -0.68 & -0.75 & -0.69 & -0.74 & -0.93 & -0.76\\
        & & $\langle \mathcal{E}_{\rm pseudo-bulge} \rangle$ & -0.49 & -0.73 & -0.82 & -0.70 & -0.66 & -0.74 & -0.68 & -0.72 & -0.93 & -0.72\\
        & & $\langle \mathcal{E}_{\rm halo} \rangle$ & -0.19 & -0.28 & -0.36 & -0.22 & -0.25 & -0.27 & -0.27 & -0.31 & -0.55 & -0.28\\ 
        \hline
        \multirow{8}{*}{\rotatebox[origin=c]{90}{Bar}} & \multicolumn{2}{c|}{Barred} & \ding{53} & \ding{53} & \checkmark & \checkmark & \ding{53} & \checkmark & \ding{53} & \checkmark & \checkmark & \checkmark & \dots{}\\
        \cline{2-14}
        & \multirow{2}{*}{Size} & $R_{\Phi}$ & \ding{53} & \ding{53} & 1.56 & 3.18 & \ding{53} & 1.94 & \ding{53} & 5.33 & 1.52 & 3.03 & \multirow{2}{*}{\dots{}}\\
        & & $R_{\rm peak}$ & \ding{53} & \ding{53} & 1.39 & 1.97 & \ding{53} & 1.64 & \ding{53} & 2.24 & 1.41 & 2.83\\
        \cline{2-14}
        & \multirow{2}{*}{Strength} & $A_{2, \rm max}(R)$ & \ding{53} & \ding{53} & 0.24 & 0.50 & \ding{53} & 0.39 & \ding{53} & 0.50 & 0.34 & 0.29 & \multirow{2}{*}{\dots{}}\\
        & & $A_{2, \rm max}(<\!R)$ & \ding{53} & \ding{53} & 0.15 & 0.32 & \ding{53} & 0.24 & \ding{53} & 0.34 & 0.25 & 0.15\\
        \cline{2-14}
        & \multirow{3}{*}{Quality flags} & $k_{\rm rot}$ & 0.32 & 0.50 & 0.43 & 0.44 & 0.38 & 0.46 & 0.39 & 0.46 & 0.70 & 0.55 & \multirow{3}{*}{\dots{}}\\
        & & $\sigma_{z}/\sigma_{R}$ & 0.77 & \ding{53} & 0.75 & 0.73 & \ding{53} & 0.73 & \ding{53} & 0.64 & 0.71 & 0.69\\
        & & $M_{b1}/M_{b2}$ & 1.01 & \ding{53} & 1.00 & 1.01 & \ding{53} & 0.96 & \ding{53} & 1.09 & 0.94 & 1.06\\
        \end{tabular}
    \label{tab:catalog_example}
\end{table*}


\subsection{Morphological classification of TNG50 \textit{z=}0 galaxies and comparison with existing catalogues}

\begin{figure*}
    \centering
    \includegraphics[width=0.83\textwidth]{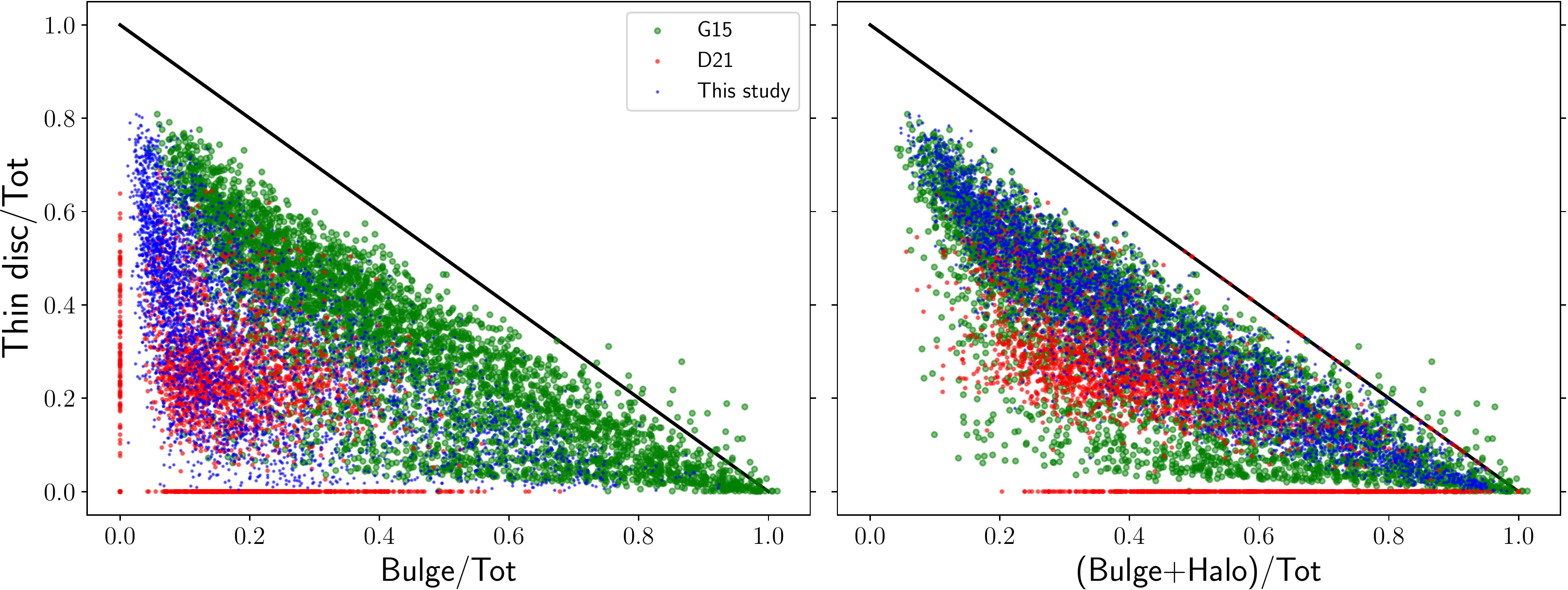}
    \caption{Mass of the thin disc component versus the bulge mass (\textit{left}) and the total spheroid mass (bulge and halo; \textit{right}).
    Our analysis (blue points) is compared with the results by \citetalias{Genel_et_al_2015} (green points) and \citetalias{Du_et_al_2021} (red points) at $z=0$. Note that for \citetalias{Genel_et_al_2015}, the single disc component is shown.
    The diagonal black line shows the $x=1-y$ trend.}
    \label{fig:disc_tot_spheroid_z0}
\end{figure*}

Fig.~\ref{fig:disc_tot_spheroid_z0} shows the ratio $M_{\rm thin}/M_{*}$ versus the ratio $M_{\rm bulge}/M_{*}$ (left panel) and the ratio $(M_{\rm bulge}+M_{\rm halo})/M_{*}$ (right panel) for every galaxy of the TNG50 run at $z=0$. 
The decomposition methods by \citetalias{Genel_et_al_2015} (green points) and \citetalias{Du_et_al_2021} (red points) are compared against the result of our decomposition (blue points).

\citetalias{Genel_et_al_2015} data have been derived by decomposing each galaxy into a thin disc and a bulge only, neglecting the remaining particles. 
The bulge is determined by doubling the mass of all the counter-rotating ($\eta<0$) particles, whereas the disc is defined as the sum of highly-rotating ($\eta>0.7$) stellar particles (which is the same threshold we apply in our method), minus the fraction of stars with $\eta<-0.7$.
%
Since the analysis by \citetalias{Genel_et_al_2015} results in a simple dichotomous decomposition (i.e. there is no distinction between bulge and halo), green points occupy the same position in both panels of Fig.~\ref{fig:disc_tot_spheroid_z0}.

On the other hand, the \citetalias{Du_et_al_2021} analysis provides a more complex outcome, where the same five morphological components are directly comparable with our results.
The similarity is further increased since we apply a very similar exclusion criterion -- as the one adopted in \citetalias{Du_et_al_2021} -- to unbound particles and/or to particles with an excessive perpendicular or parallel angular momentum (see \S~\ref{subsec:decomp_alg} for details).
Unfortunately, the $z=0$ sample of \citetalias{Du_et_al_2021} is far less numerous than \citetalias{Genel_et_al_2015}'s and ours, since they performed the analysis only on those galaxies with $M_{*}>10^{9}~\msun$, whereas our threshold is based on a minimum number of $10^4$ stellar particles, which is satisfied by 4091 galaxies at $z=0$.
We further note that \citetalias{Genel_et_al_2015} uses an even lower mass threshold, corresponding to $3.4 \times 10^{8}~\msun$, computed within $2r_{\rm hm}$.

As we see, a sub-sample ($<1$~percent) of \citetalias{Genel_et_al_2015} galaxies with low ratio $M_{\rm thin}/M_{\rm spheroid}$ (where $M_{\rm spheroid}$ is the combined mass of the stellar bulge and halo) shows values larger than unity.
For these objects bulge masses are likely to be overestimated.\footnote{The issue can be due to the fact that the morphological components are not strictly identified from their constituent particles, but are only estimated from smaller intervals of the circularity distribution.}
This behavior is not observed in our and \citetalias{Du_et_al_2021}'s data, where $M_{*} \geq M_{\rm thin} + M_{\rm thick} + M_{\rm pseudo-bulge} + M_{\rm bulge} + M_{\rm halo}$ (this is not an equality relation because of the excluded unbound particles, see above).

\begin{figure}
    \centering
    \includegraphics[width=0.47\textwidth]{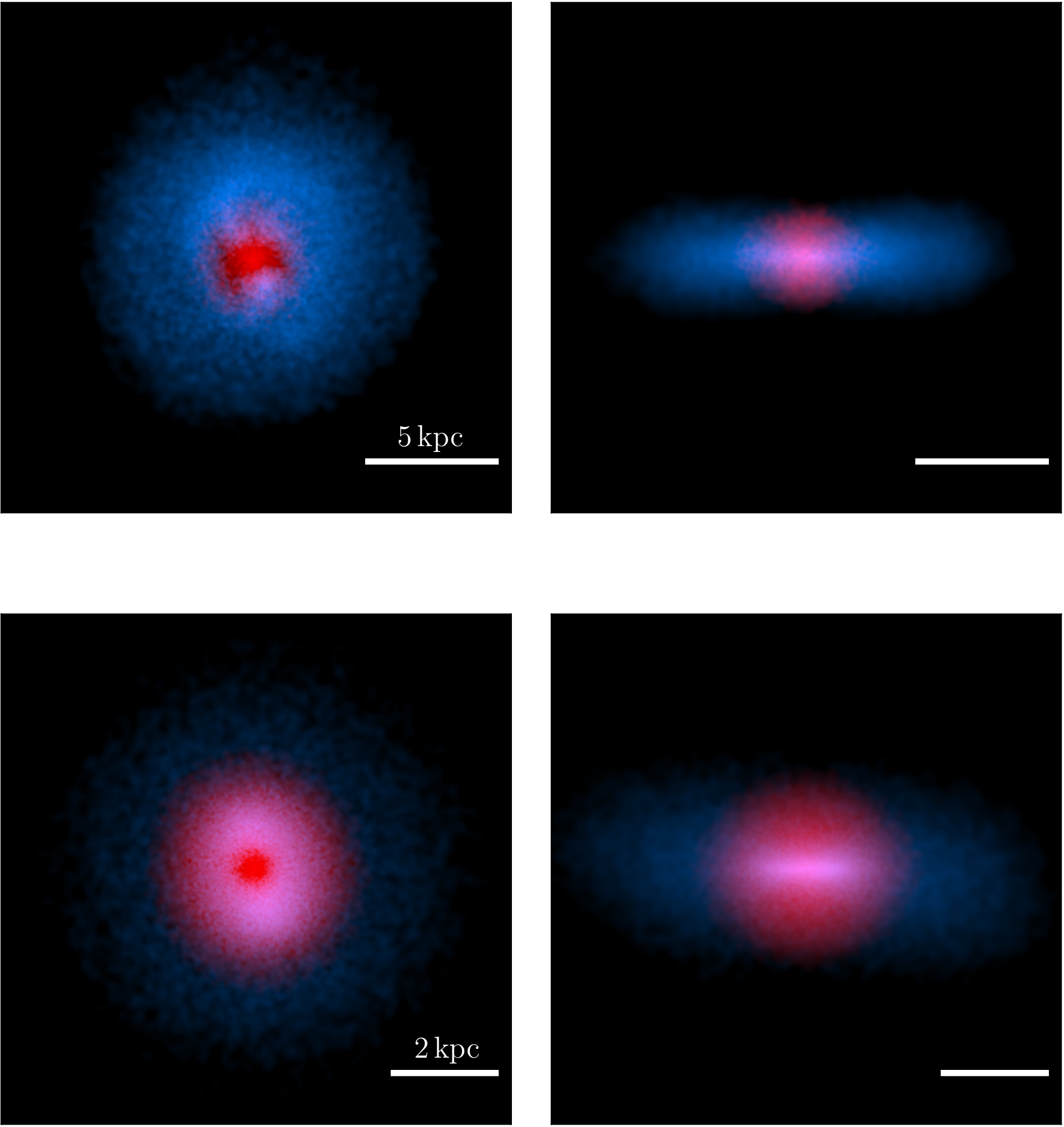}\\
    \caption{RGB composite images of two $z=0$ galaxies: a disc galaxy in the \textit{top row} (ID 630870) and a spheroidal one in the \textit{bottom row} (ID 338447).
    We show in blue the thin and thick disc components, in red the bulge, and in white the remaining stars.
    Densities range from $10^{7}$, to $5\times10^{9}~\msun$~pc$^{-2}$ for the disc galaxy and from $5\times10^{7}$, to $5\times10^{10}~\msun$~pc$^{-2}$ in the case of the spheroidal system.}
    \label{fig:high_components}
\end{figure}

In $32$~percent of objects \citetalias{Du_et_al_2021} do not identify either a bulge or a thin disc (or the two of them), whereas our method always recognizes at least a combination of these two stellar structures.\footnote{Notice that, although we do not find any outlier with either $M_{\rm thin}=0$ or $M_{\rm bulge}=0$ among the $z=0$ galaxies of TNG50, our algorithm does not enforce any component to be present in the decomposition.}
As an example, in Fig.~\ref{fig:high_components} we show a disc galaxy (top row) and a spheroidal galaxy (bottom row) at $z=0$, where we highlight the thin and thick disc components (blue) and the bulge (red) and where \citetalias{Du_et_al_2021} decomposition finds $M_{\rm thin} + M_{\rm thick} = 0$ and $M_{\rm bulge} = 0$ in the first and second system, respectively.
On the contrary, our method associates $65$~percent of the mass to the disc in the top galaxy and $54$~percent of the mass to the bulge in the bottom one.

\begin{figure*}
    \centering
    \includegraphics[width=0.83\textwidth]{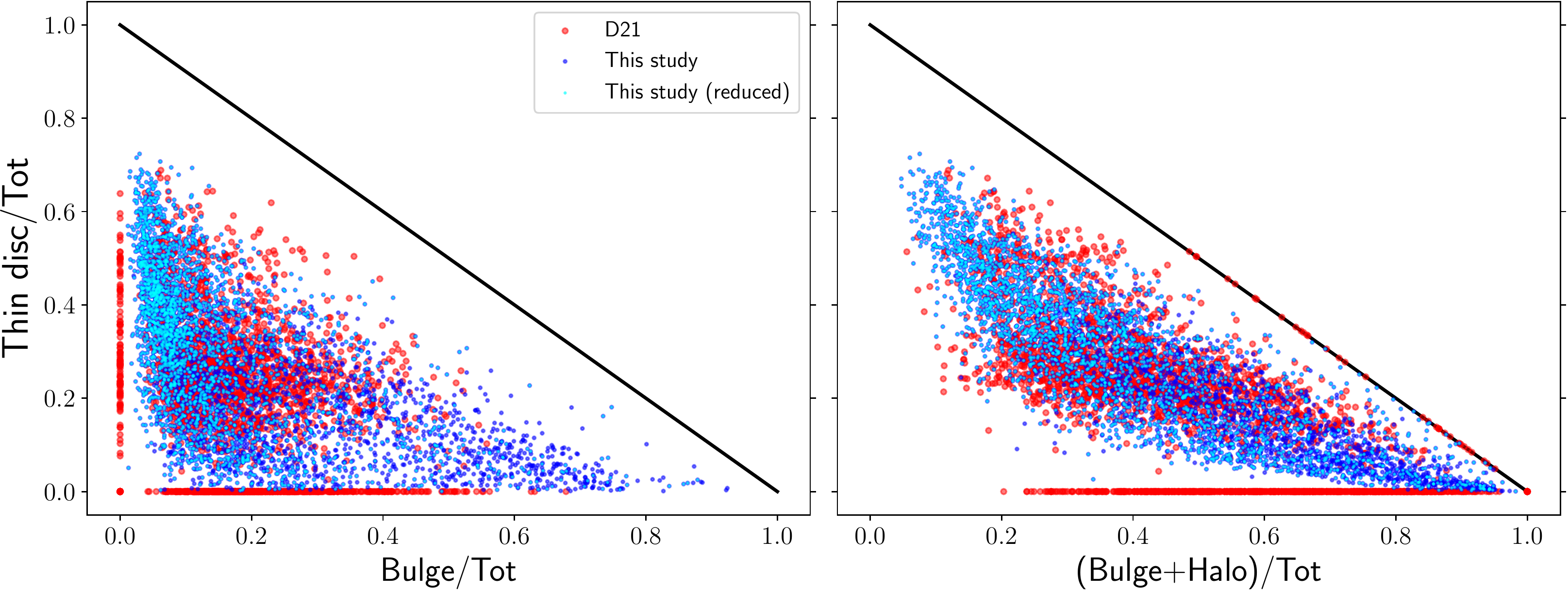}
    \caption{Same comparison of Fig.~\ref{fig:disc_tot_spheroid_z0} between \citetalias{Du_et_al_2021}'s data (red points) and our method, this time using a circularity threshold $\eta=0.8$ to define the thin disc component (blue points).
    We also show a reduced sample matching one-to-one the galaxies analysed by \citetalias{Du_et_al_2021} (cyan points).}
    \label{fig:disc_tot_spheroid_z0_j0.8}
\end{figure*}

Aside from the numerous similarities and a general good agreement with the two other methods, we note some interesting differences, in addition to the outliers previously discussed (such as the cumulative mass exceeding unity for \citetalias{Genel_et_al_2015}, or the bulgeless/discless galaxies of \citetalias{Du_et_al_2021}). 
From left panel of Fig.~\ref{fig:disc_tot_spheroid_z0}, it emerges that our method always yields an $M_{\rm thin}/M_{\rm bulge}$ ratio lower than that of \citetalias{Genel_et_al_2015}.
As already discussed, this difference almost disappears in the right panel, since \citetalias{Genel_et_al_2015} does not disentangle the different types of spheroids.
Moreover, neither our sample, nor \citetalias{Du_et_al_2021}'s one show the large amount of low-mass discs visible in the \citetalias{Genel_et_al_2015} distribution at all spheroid mass values.
In general, our catalogue shows galaxies with a higher $M_{\rm thin}/M_{*}$ ratio with respect to \citetalias{Du_et_al_2021}. While it is hard to see this trend in the left panel of Fig.~\ref{fig:disc_tot_spheroid_z0}, probably because of the different numbers of objects analysed, the difference becomes clearer in the right panel.
This discrepancy could be due to the higher threshold the authors adopt to identify the thin (cold) disc, i.e. they select, as a thin disc, the 3D Gaussians in the space $\eta-j_{p}/j_{\rm circ}-\mathcal{E}$ with a mean circularity $\langle \eta \rangle>0.85$.
To test this possibility, in Fig.~\ref{fig:disc_tot_spheroid_z0_j0.8} we show the outcome of our procedure if we increase our circularity threshold for the thin disc component to $\eta>0.8$.
As a further test, we also show as cyan points a sub-sample of our complete dataset matching the sample by \citetalias{Du_et_al_2021}, composed only by unbarred galaxies.
Clearly, the agreement between the methods improves for the disc mass distribution, whereas some peculiar dissimilarities in the spheroids remain.
\citetalias{Du_et_al_2021}'s galaxies have more massive bulges with respect to ours:
red points in the left panel of Fig.~\ref{fig:disc_tot_spheroid_z0_j0.8} have a median of $M_{\rm bulge}/M_{*}=0.16$, whereas our method produces a median of $M_{\rm bulge}/M_{*}=0.14$ on the whole population (blue points), and even $M_{\rm bulge}/M_{*}=0.10$ in the reduced sample (cyan points).
If we consider the total spheroid mass $M_{\rm spheroid}$ (right panel) instead, our algorithm produces a more uniformly distributed sample with respect to \citetalias{Du_et_al_2021}.
Interestingly, we find only small differences between our full sample and the reduced one, in which bulges (and halos) are slightly under-massive in the second case.
To summarize, our code generally finds slightly less massive spheroids, and more massive discs. 

\begin{figure*}
    \centering
    \includegraphics[width=0.83\textwidth]{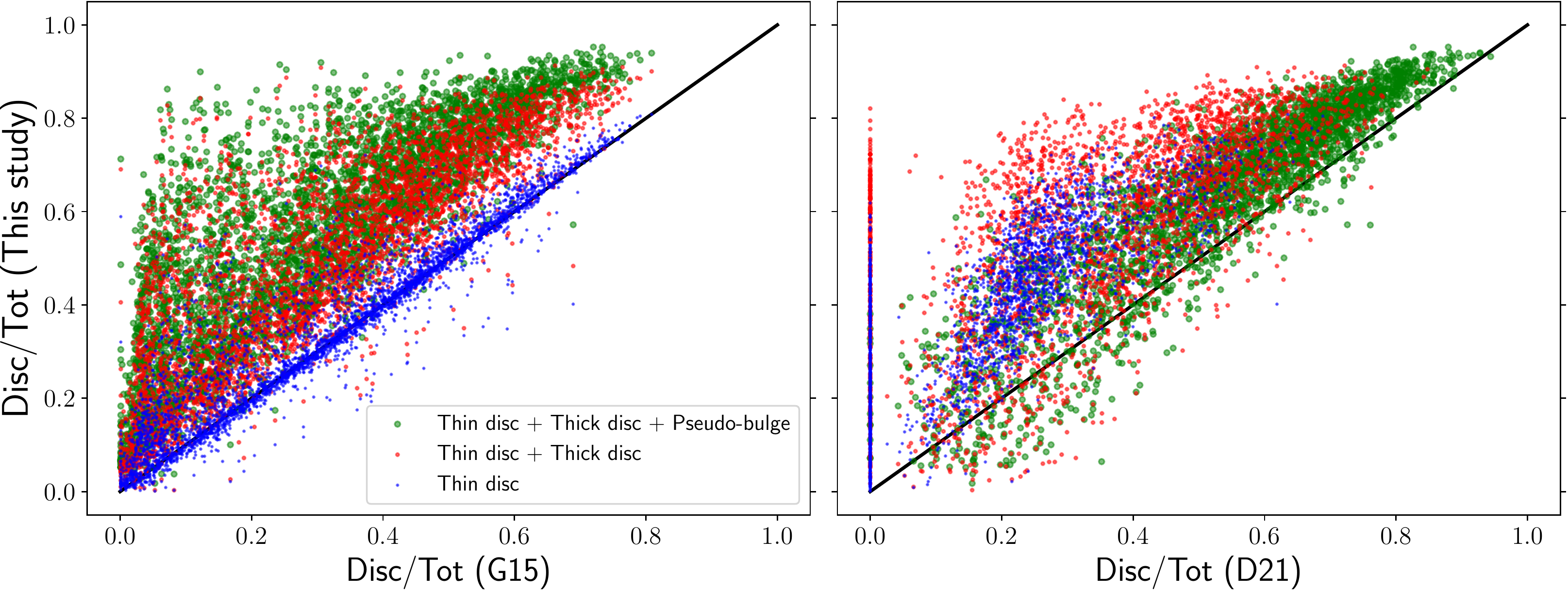}
    \caption{The results of our decomposition are compared against the catalogue by \citetalias{Genel_et_al_2015} ({\it left}) and \citetalias{Du_et_al_2021} ({\it right}) for the thin disc only (blue points), thin and thick disc components (red points), and for the whole disc-like components (including pseudo-bulge, green points).}
    \label{fig:disc_tot_z0}
\end{figure*}

The importance of the whole disc structure, as it is determined from the different techniques, is compared in Fig.~\ref{fig:disc_tot_z0}.
The better agreement between our results and \citetalias{Genel_et_al_2015}'s decomposition surely lies on the thin disc component, given the very similar definition as we already discussed.
The deviations from the line $y=x$, especially for $M_{\rm thin}/M_{*}<0.2$, can be due to inaccuracies in the rotation procedure of the galaxy which can lead to underestimates in the perpendicular component of the angular momentum.
Obviously, because of our definition of ``disc-like component'', the disc mass (almost) monotonically increases as the other morphological systems are added up.
Since in \citetalias{Genel_et_al_2015} the thick disc and the pseudo-bulge are not considered at all,
there are objects that can have a rotating component of $\sim 90$~percent in our catalogue, and only $\sim 10$~percent according to \citetalias{Genel_et_al_2015}.\footnote{There are also a few objects ($\sim0.3$~percent) where even the total disc mass is higher in \citetalias{Genel_et_al_2015}'s decomposition, likely because of a different assignment procedure of the star particles to the spheroidal components.}
This aspect may result in some serious consequences, particularly when an analysis is performed only on a specific morphological class of galaxies (e.g., only disc galaxies).
We will further discuss this topic in the peculiar case of barred galaxies in Section~\ref{subsec:bar_structures}.
Analogously to the cyan dataset of Fig.~\ref{fig:disc_tot_spheroid_z0_j0.8}, in the right panel of Fig.~\ref{fig:disc_tot_z0} we are forced to reduce our sample in order to match \citetalias{Du_et_al_2021}'s galaxies.
As previously seen, we generally find more massive discs (and less massive spheroids) with respect to \citetalias{Du_et_al_2021}.
Remarkably, among all the \citetalias{Du_et_al_2021}'s outliers of Fig.~\ref{fig:disc_tot_spheroid_z0} with $M_{\rm thin}=0$, numerous galaxies have also $M_{\rm thick}=0$.
The degeneracy is only broken by the pseudo-bulge component, which reaches up to $\sim 80$~percent of the galactic stellar mass in some cases.
In the light of this, it seems that our code does produce more realistic results, where pseudo-bulges never play such a major role in the galaxy kinematics.
Moreover, it is again clear that, despite our thin discs being almost always more dominant than \citetalias{Du_et_al_2021}'s cold discs, the agreement significantly improves when the total disc mass is considered (green points) and is nevertheless always better than in the \citetalias{Genel_et_al_2015} case.

\begin{figure}
    \centering
    \includegraphics[width=0.5\textwidth]{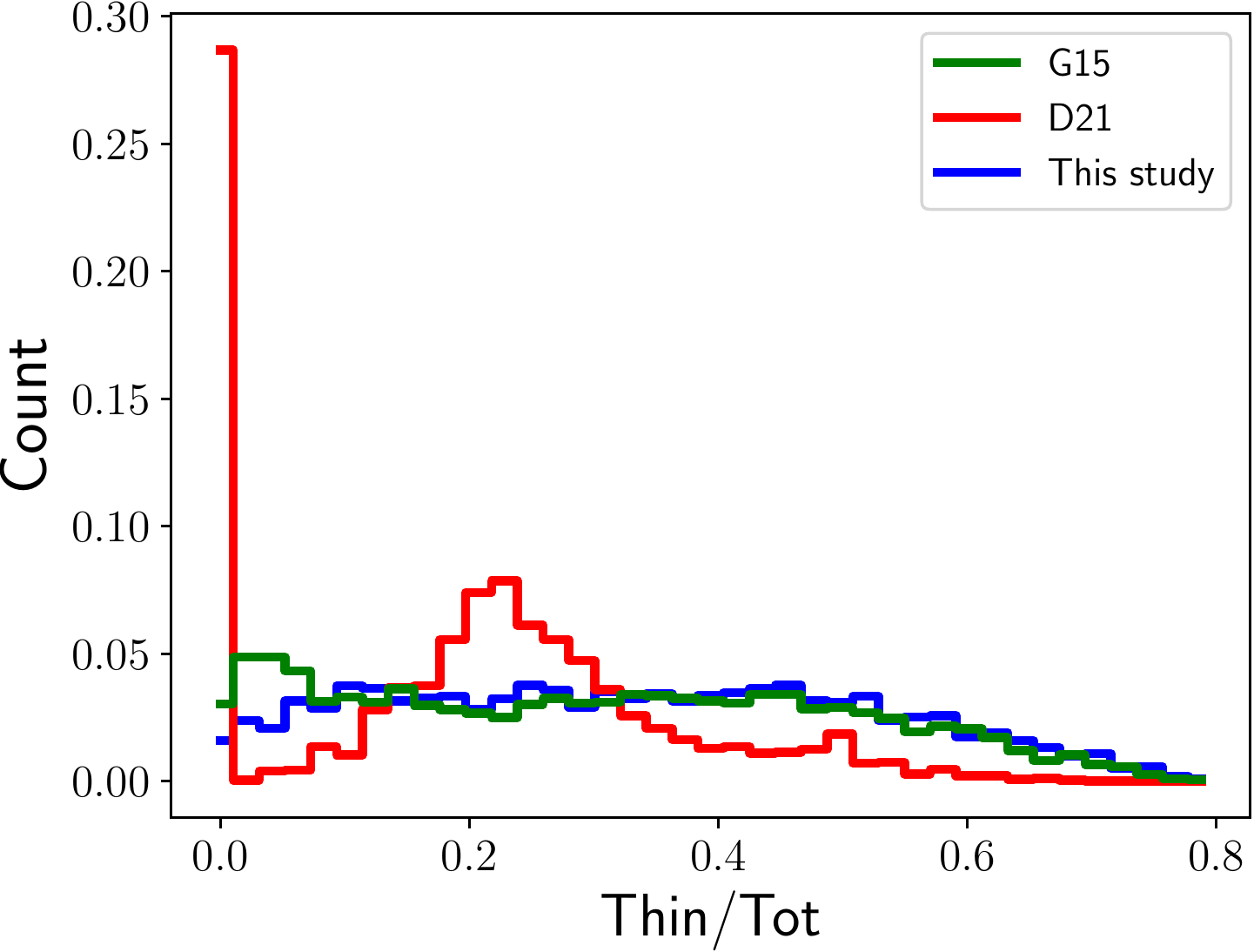}
    \caption{Distribution of $z=0$ galaxies according to their $M_{\rm thin}/M_{*}$ ratio. The methods by \citetalias{Genel_et_al_2015} (green) and \citetalias{Du_et_al_2021} (red) are shown along with our results (blue).}
    \label{fig:Histo_count_thin_z0}
\end{figure}

To summarize, in Fig.~\ref{fig:Histo_count_thin_z0} we compare the distribution of the thin disc mass fraction of the whole $z=0$ galaxy population among the different catalogues.
The distribution is fairly constant in our sample, with two slowly decreasing tails for $M_{\rm thin}/M_{*} \simeq 0.1$ and $M_{\rm thin}/M_{*} \gtrsim 0.6$.
Apart from the low-mass disc tail,
our method agrees well with \citetalias{Genel_et_al_2015}'s results, when considering only this single stellar component.
On the other hand, the \citetalias{Du_et_al_2021} catalogue shows a peculiar and discontinuous distribution, with numerous galaxies having under-massive or absent thin disc components and a peak around $M_{\rm thin}/M_{*} \simeq 0.3$.
This may be due to the fact that \citetalias{Du_et_al_2021} algorithm might not identify a component when it is poorly populated, rather merging it to the closest Gaussian in the phase space.

\begin{figure}
    \centering
    \includegraphics[width=0.5\textwidth]{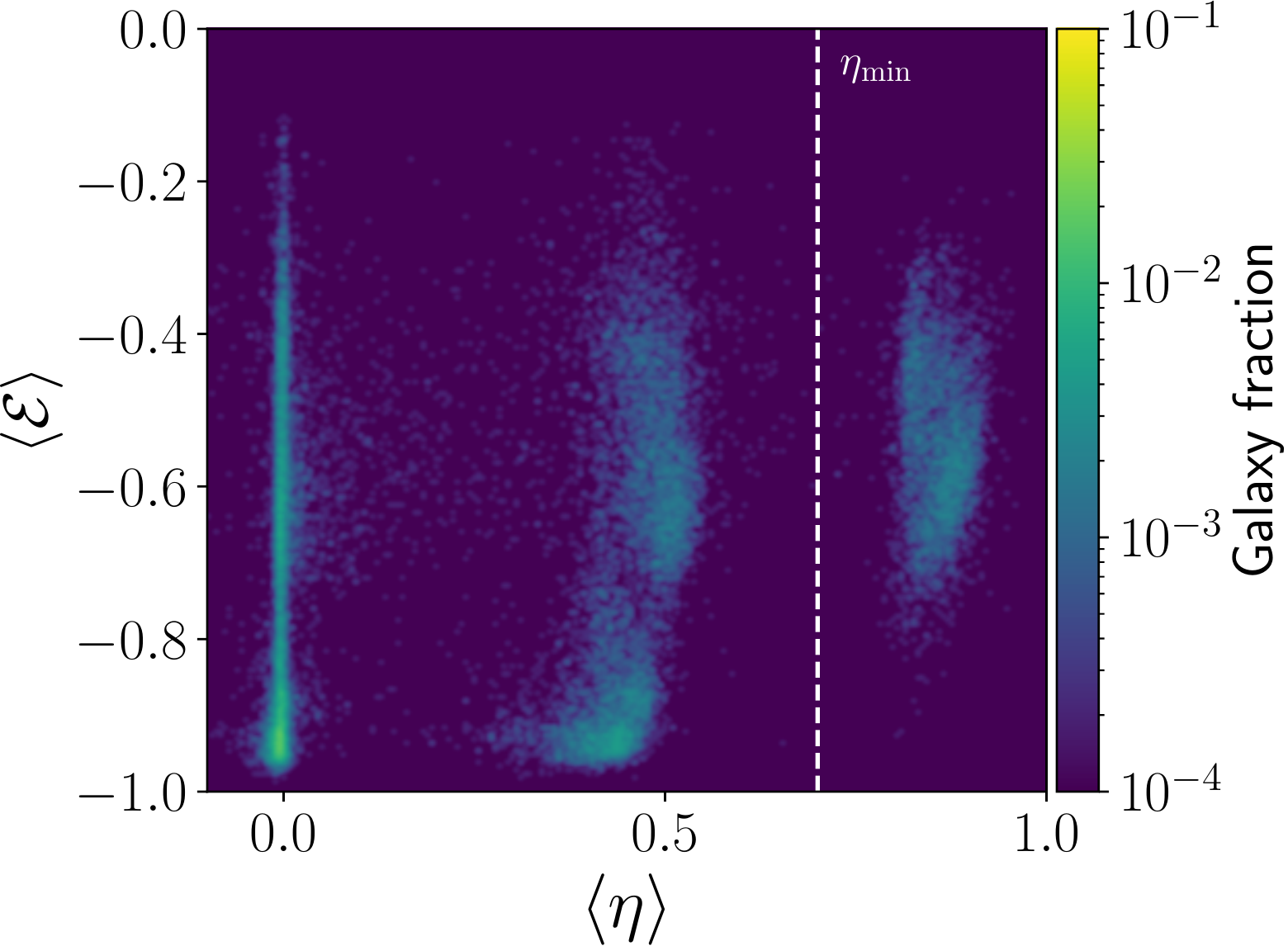}
    \caption{Kinematic phase space distribution of all the morphological components we identify for the $z=0$ TNG50 galaxies.
    Each component is represented by its mean circularity $\eta$ and mean energy $\mathcal{E}$.
    The vertical dashed line marks the threshold of $\eta=0.7$, here adopted to define the thin disc part.}
    \label{fig:z0_EvsJ}
\end{figure}

Our complete sample at $z=0$ is shown inside the kinematic phase space ($\eta-\mathcal{E}$) in Fig.~\ref{fig:z0_EvsJ}.
Each galaxy component is represented through its mean circularity $\langle \eta \rangle$ and mean total energy $\langle \mathcal{E} \rangle$.
The five morphological components cluster in five recognizable regions in the $j-E$ plane, more or less separated.
The median circularities for these regions are 0.86, 0.49, 0.43, 0, and 0 for the thin disc, thick disc, pseudo-bulge, bulge, and halo, respectively.
The median value for the spheroidal components shows the success of our method in identifying non-rotationally-supported systems.
A natural energy boundary seems to lie around $\langle \mathcal{E} \rangle = -0.7$, although not particularly evident.
Since our algorithm determines $E_{\rm cut}$ in a completely unsupervised fashion, without requiring any guess or a priori choice, this result seems to be intrinsically related to the galaxy population.

\subsection{Redshift evolution}
\label{subsec:redshift_evolution}

Here we take advantage of the flexibility of our method, which allows us to reliably and efficiently decompose all the snapshots of TNG50 and give some hints about the evolution of galaxy morphologies.
Specifically, we do not refer to the evolution of individual galaxies, as we do not trace back their progenitors. We instead discuss how the relative importance of the different components globally change with redshift.

\begin{figure*}
    \centering
    \includegraphics[width=0.93\textwidth]{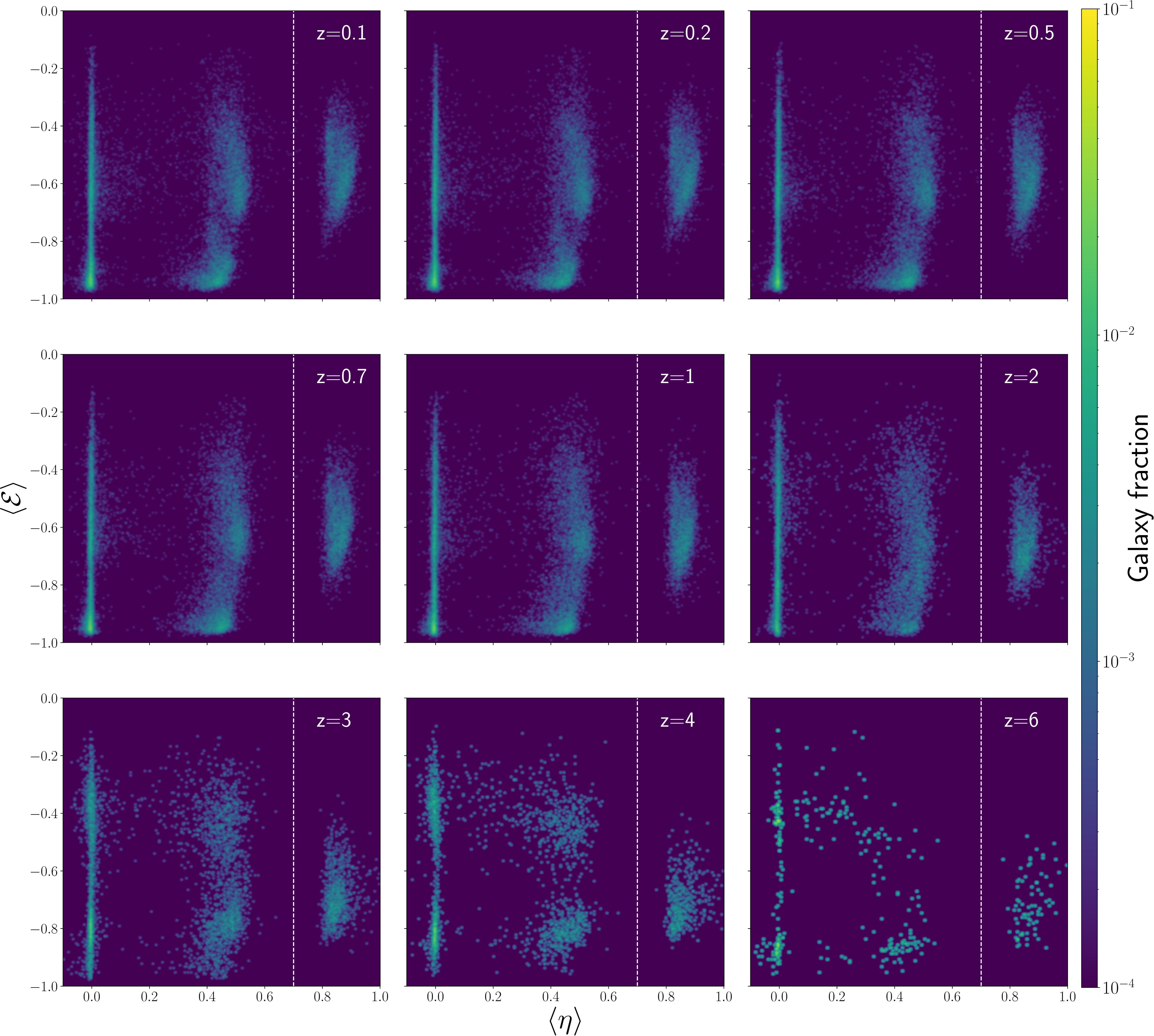}
    \caption{Redshift evolution of the phase space distribution: the same representation of Fig.~\ref{fig:z0_EvsJ} is presented again for a sample of redshifts from the TNG50 simulation, namely $z=0.1,0.2,0.5,0.7,1,2,3,4,6$.}
    \label{fig:EvsJ}
\end{figure*}

Fig.~\ref{fig:EvsJ} extends the phase space representation of Fig.~\ref{fig:z0_EvsJ} to a sub-sample of snapshots from $z=0.1$ to $z=6$.
From Fig.~\ref{fig:z0_EvsJ} and \ref{fig:EvsJ}, it seems that there is no clear evolution in the cumulative phase space in the redshift range $z=[0;1]$.
This does not mean that galaxies are not evolving, but simply that, in general, the binding energies and circularities of the main kinematic components are preserved as in the last snapshots of the simulations.
At $z=2$, the thick disc and the pseudo-bulge clouds overlap, hinting to a fast evolution and a less clear boundary between the mildly rotating components.
Probably due to the numerous mergers and dynamical events occurring around $z=2$, the various morphological constituents show no clear separation in energy.
A clear energy separation is again visible for $z\geq3$, although this is mostly due to the much lower number of objects.
For the same reason, also the spheroidal components (bulge and halo) are easily recognizable only at $z>4$, since they occupy a contiguous energy range centred at $\langle \eta=0 \rangle$ for the rest of the simulation time.
%
Interestingly, the thick disc is the component showing the largest changes over the entire redshift interval, both in binding energy and in circularity.
While the spread in $\langle \eta \rangle$ increases at higher redshift, where galaxies are still forming, continuously interacting and violently accreting material from the environment, the mean $\eta_{\rm thick}$ almost monotonically moves toward higher values with time ($\langle \eta_{\rm thick} \rangle$ ranges from 0.24 at $z=6$, to 0.48 at $z=0$).
The mean $\mathcal{E}_{\rm thick}$ shows a similar evolution to a more bound state with lower values ($\langle \mathcal{E}_{\rm thick} \rangle=-0.42$ at $z=6$, and $\langle \mathcal{E}_{\rm thick} \rangle =-0.53$ at $z=0$).
This evolution in the mildly rotating component is consistent with the picture of galactic discs being thicker and more turbulent at higher redshift \citep[see, e.g.,][]{Kassin_et_al_2012} because of the more frequent dynamical interactions and the higher star formation activity (which, in turns, results in stronger feedback from supernovae), and their later settling in a quieter and thinner distribution with time.

\begin{figure*}
    \centering
    \includegraphics[width=0.83\textwidth]{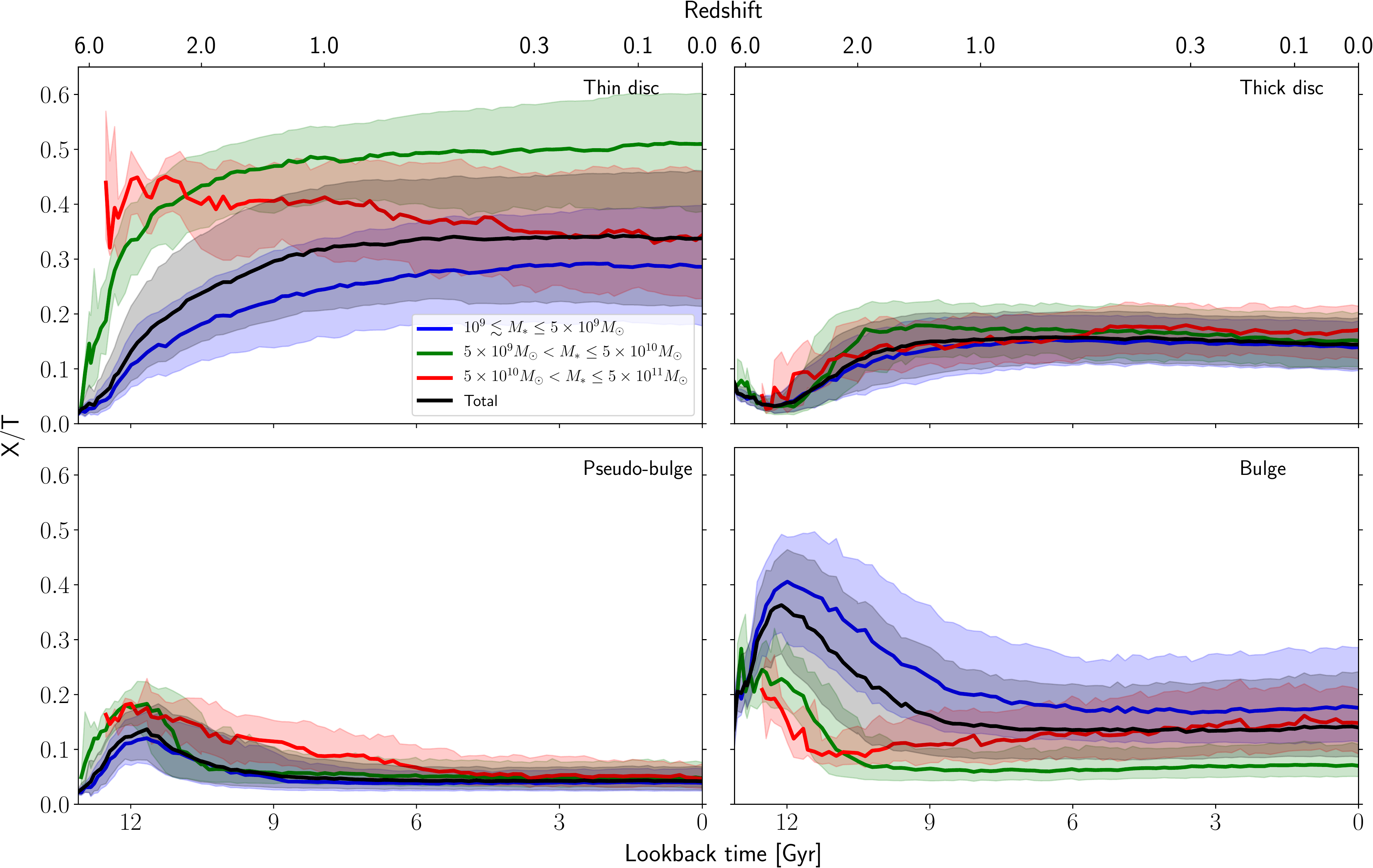}\\
    \caption{Redshift evolution of the median mass fractions of the single morphological components.
    Clockwise from top left: thin disc, thick disc, bulge, and pseudo-bulge in three different mass bins, namely $10^{9} \lesssim M_{*}/\msun \leq 5 \times 10^{9}$ (blue lines), $5 \times 10^{9} < M_{*}/\msun \leq 5 \times 10^{10}$ (green lines), and  $5 \times 10^{10} < M_{*}/\msun \leq 5\times10^{11}$ (red lines).
    Black lines show the trend for the whole population in the mass range $10^{9}\lesssim M_{*}/\msun < 6\times10^{12}$.
    Shaded areas extend from the $30$th, to the $70$th percentile of each distribution.}
    \label{fig:component_fraction}
\end{figure*}

\begin{figure}
    \centering
    \includegraphics[width=0.45\textwidth]{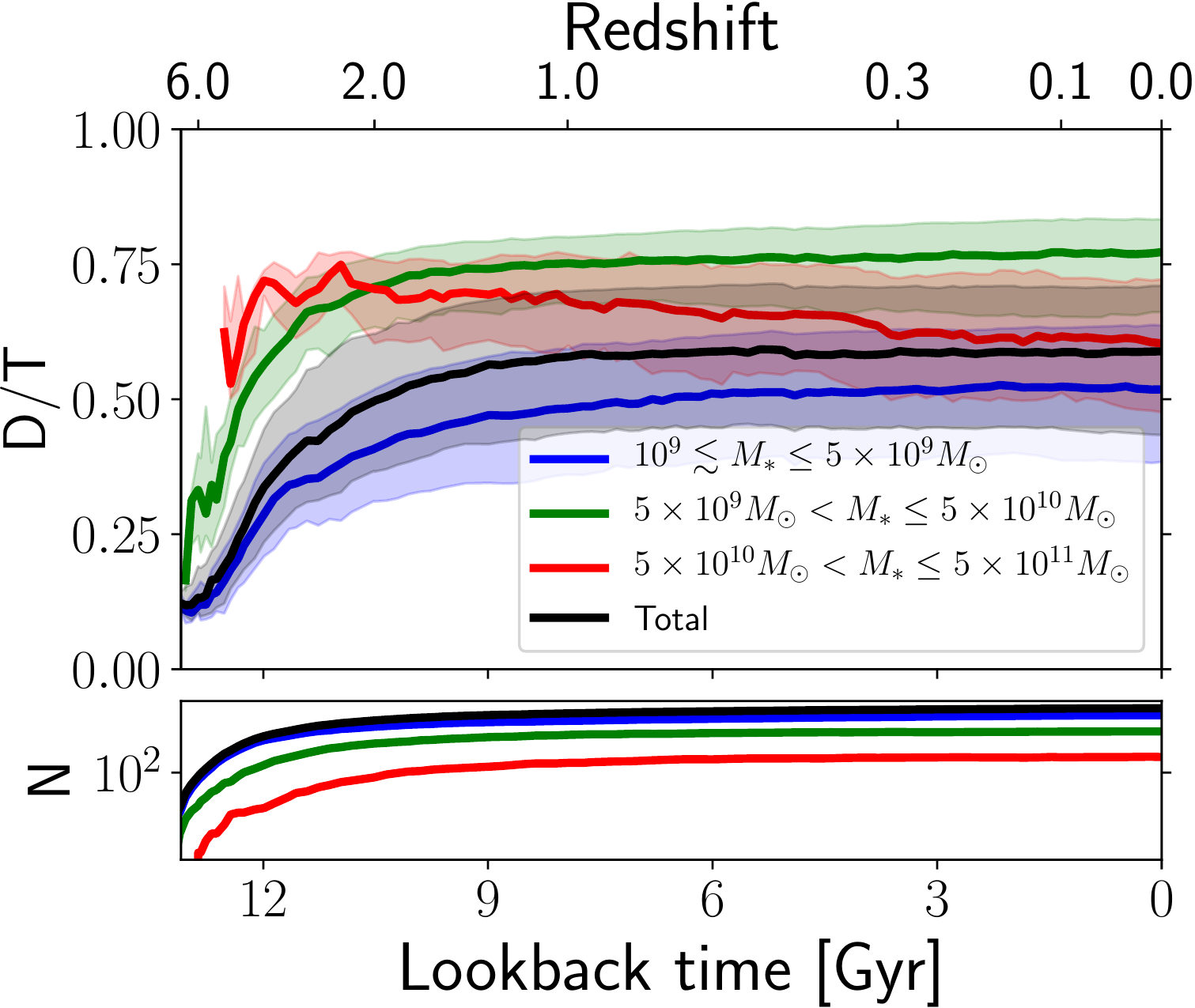}\\
    \caption{{\it Top panel}: redshift evolution of the median fraction $M_{\rm disc}/M_{*}$ (where the disc mass $M_{\rm disc}$ here is the sum of all the rotating components), for the same mass bins used in Fig.~\ref{fig:component_fraction} and with the same colour code.
    Shaded regions show the ranges between the 30th and the 70th percentile of each distribution.
    {\it Bottom panel}: total number of galaxies in the different mass bins.}
    \label{fig:D_T}
\end{figure}

On the contrary, the dynamical processes responsible for the building-up of the most massive elliptical galaxies we observe in the local Universe (i.e. merger events which lower the circularity and increase the random motion of stars) clearly play a minor role in the analysed sample, and take place only in a handful of (massive) objects, giving the relatively small simulation box, which does not claim to be a statistically representative volume of the real Universe.
This signature can be more easily seen in Fig.~\ref{fig:component_fraction} where we show, for the thin disc, the thick disc, the pseudo-bulge, and the bulge, the evolution with redshift of the median mass fraction in different bins of galactic stellar mass.
In the following, we divide the sample of objects into three mass bins, selected at each redshift, i.e. $10^{9} \lesssim M_{*}/\msun \leq 5 \times 10^{9}$ (where the lower limit comes from the threshold on the minimum number of star particles; see \S~\ref{sec:method}), $5 \times 10^{9} < M_{*}/\msun \leq 5 \times 10^{10}$ (a Milky-Way like mass interval), and $5 \times 10^{10} < M_{*}/\msun  \leq 5\times10^{11}$, chosen in order to guarantee a statistically representative and significant\footnote{In the following figures a mass bin is shown at its corresponding redshift $z$ only if the number of galaxies at that redshift minus the Poissonian error is larger than zero, i.e. $N_{\rm gal}(z)-\sqrt{N_{\rm gal}(z)}>0$.} number of galaxies in each bin (the number of objects in each mass bin is visible in the lower panel of Fig.~\ref{fig:D_T}).
\noindent The TNG50 simulation leads also to the formation of a few massive galaxies, with $5 \times 10^{11} < M_{*}/\msun < 6\times10^{12}$, that we do not include in the most massive mass bin, for statistical significance.
In the total sample, with no mass cut in the range $10^{9}\lesssim M_{*}/\msun < 6\times10^{12}$ (black lines), the pseudo-bulge component reaches a maximum at about $z=5$ and then drops in favour of the disc components, which increase their importance as galaxies grow and maintain their values from $z\sim1$ until $z=0$.
This trend is visible both in the lower and intermediate mass bins, though with some differences among the components.
The ratio $M_{\rm thin}/M_{*}$ is always the lowest for the smaller-mass galaxies, whereas the ratio $M_{\rm bulge}/M_{*}$ is always the highest: the smaller galaxies are generally irregular and turbulent, especially at higher redshift, thus their motion is dominated by a non-rotating component.
These systems switch from being dispersion-dominated (large bulge components), to thin disc-dominated after $z\sim2$.

As anticipated, the highest mass bin shows the most noticeable evolution after $z=2$: massive spheroids hosted at the centre of the highest dark matter over-densities form likely through various major mergers.
These events are believed to destroy rotationally supported systems by changing the overall potential well and, via violent relaxation, yield systems supported by random motion.
The red line in the top-left panel of Fig.~\ref{fig:component_fraction}, the thin disc, is the only one that decreases monotonically toward $z=0$, while the bulge (bottom-right panel) of most massive galaxies is the only one constantly increasing its importance over $\approx 10$~Gyr, eventually reaching the median of the entire population.
Curiously, the evolution of the thick disc and pseudo-bulge components seems independent of the galactic mass.
Only a small difference is visible in the pseudo-bulge of most massive galaxies, where its relative contribution decreases slower with respect to other mass bins.
Fluctuations at high-$z$ in the most massive objects are due to the low statistics of the subsets.

It is absolutely non trivial to provide a reliable kinematic definition of a disc galaxy, in order to mimic the Hubble morphological classification.
A ``disc galaxy'' is a conventional construct to define a system where the majority of stellar orbits have an aligned angular momentum, but it strongly depends on the appearance of that system.
In real galaxies there is a continuous spectrum of morphologies and the selection of a threshold is inevitably arbitrary.
This definition can have serious consequences, for instance, in the analysis of the output of cosmological simulations, where the studies have to be narrowed down to e.g. disc galaxies only, in order to study the formation of sub-structures.
In this work we check the consistency of two definitions based on our kinematic decomposition method.
In particular, we define a disc galaxy as either (i) a system in which the cumulative mass of its rotating components accounts for the major part of the galaxy stellar mass, i.e. $M_{\rm thin}+M_{\rm thick}+M_{\rm pseudo-bulge}>0.5M_{*}$ -- our fiducial choice -- or (ii) a system with extended disc-like components dominating over more centrally concentrated structures, i.e. $M_{\rm thin}+M_{\rm thick}>M_{\rm bulge}+M_{\rm pseudo-bulge}$.
We note that it is not straightforward to disentangle pseudo-bulges from classical bulges in observations. Various techniques are used (e.g. the study of stellar kinematics, photometric profile, etc.) but none of them guarantee a unique identification \citep[see, e.g.,][and references therein]{Kormendy_Kennicutt_2004}.

\begin{figure*}
    \centering
    \includegraphics[width=0.83\textwidth]{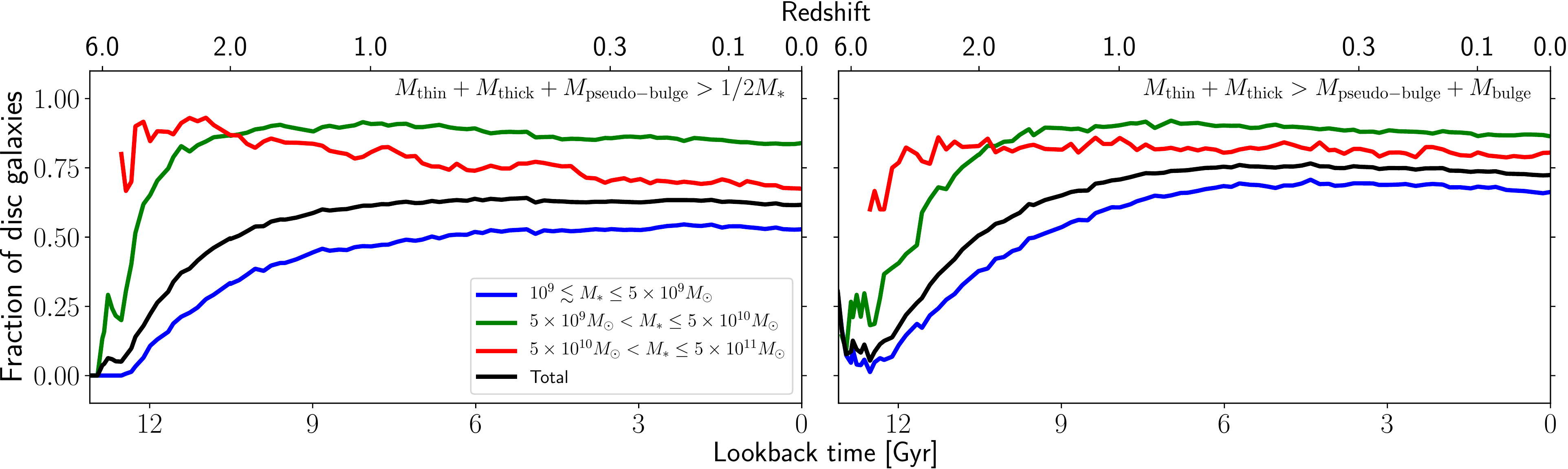}\\
    \caption{Redshift evolution of the number of ``disc galaxies'' over the total number of objects in the same mass bins, with the same colour code used in Fig.~\ref{fig:component_fraction}.
    Two definitions of ``disc galaxy'' are applied: $M_{\rm thin}+M_{\rm thick}+M_{\rm pseudo-bulge}>0.5M_{*}$ ({\it left panel}) and $M_{\rm thin}+M_{\rm thick}>M_{\rm bulge}+M_{\rm pseudo-bulge}$ ({\it right panel}).}
    \label{fig:disc_number}
\end{figure*}

In Fig.~\ref{fig:disc_number}, we show the redshift evolution of the fraction of disc galaxies over the total number of objects, according to these two definitions.
The trends are similar for both disc definitions.
In agreement with our previous findings, the disc galaxy population in the lowest mass bins becomes dominant (larger than 50~percent) after $z\sim1$ ($z\lesssim1$ in the left panel and $z\sim1.5$ in the right panel) and maintains an almost constant value until $z=0$.
In intermediate-mass galaxies discs reach almost 90~percent of the sample from about $z=2$, and show only a minor decrease toward $z=0$.
We note nonetheless that the high-$z$ ($z \gtrsim 2$) evolution of most massive galaxies should be taken with caution, because of the very low statistics.
In general, smaller system are less likely to be disc galaxies with respect to more massive objects at all redshifts and more disc galaxies can be found using the second definition of discs, even though their number grows slightly slower.
A minimal decrease in all the fractions is visible at lower redshifts, regardless of the definition adopted.
This agrees with the scenario where both bulges and pseudo-bulges grow at the expense of thin and thick discs.
The former because of major mergers between galaxies, the latter as a consequence of various secular evolutionary processes.
This evolution can also be tentatively seen in Fig.~\ref{fig:component_fraction} at $z\lesssim 0.3$, where a slight decline in the disc components is accompanied by a small increase in the bulge medians.
In conclusion, both definitions of disc galaxy give similar results on TNG50 objects.
Hence, for the purpose of this work, we hereafter define disc galaxies as those resulting from the first ``kinematic definition'' (this criterion has already been applied to select galaxies in Fig.~\ref{fig:Spiral_Elliptical_dist} and \ref{fig:Spiral_Elliptical}).

In addition to the computation of the number of disc galaxies, in Fig.~\ref{fig:D_T} we provide the redshift evolution of the median of the disc-to-total ratio, where the disc is determined according to our fiducial definition.
The median $M_{\rm disc}/M_{*}$ follows almost the same evolution of the total number of disc galaxies, reaching about $59$~percent in the cumulative sample.
The most massive galaxies show the highest ratio at $z=6$, i.e. about $75$~percent, and then decline toward $z=0$.
A galaxy with a mass $M_{*}>5\times 10^{10}\msun$ near the epoch of re-ionization likely lies at the centre of the most massive dark matter haloes in the simulation and, interestingly, has about two third of its stars orbiting with a significant angular momentum component.
We note that galaxies with $5 \times 10^{11} < M_{*}/\msun < 6 \times 10^{12}$ (here included only in the total sample) have, predictably, the lowest $M_{\rm disc}/M_{*}$ ratio, slightly above 0.25 after $z=1$.

\citet{Pillepich_et_al_2019} provided a study of the morphological evolution of TNG50 galaxies, finding ``disc galaxies'' to be more frequent at low redshift and high stellar masses, in agreement with some observational results on the CANDELS galaxies \citep{van_der_Wel_et_al_2014, Zhang_et_al_2019}.
However, a direct comparison with their data is not straightforward, since their analysis exclusively relies on the aspect ratios of stellar distribution \citep{van_der_Wel_et_al_2014} and on the kinematic probe $v_{\rm rot}/\sigma_{*}$ (where $v_{\rm rot}$ is the peak of the galaxy rotation curve, and $\sigma_{*}$ the mean value of the galaxy velocity dispersion) with no distinction among the possible various galaxy sub-structures.
Nevertheless, we still find a compatible growth of the disc galaxy fraction with respect to the lookback time, that reaches higher values at higher masses until $M_{*}=5\times10^{10}~\msun$.
On the contrary, we notice a less evident evolution after $z=2$ and observe a slightly opposite trend for $M_{*}>5\times10^{10}~\msun$.
%

\subsection{Bar structures}
\label{subsec:bar_structures}

%
The definition of ``bar fraction'' is not unique, but depends on the selected sample.
At $z=0$, if we consider only the disc galaxies according to our first definition, we would find only 433 (633) barred galaxies with $A_{2, \rm max}(R)\geq0.2$ ($A_{2, \rm max}(R)\geq0.1$), against the total number of barred galaxies of 520 (770), with no constraint imposed on the galaxy morphology.
As an example, the top panel of Fig.~\ref{fig:bar_ex} shows the stellar density map of a strongly barred galaxy ($A_{2, \rm max}(R)=0.59$ and $R_{\Phi}=3.17$~kpc) identified as a spheroidal galaxy, with $M_{\rm bulge}$ and $M_{\rm halo}$ combined to be more than $80$~percent of the total stellar mass of the system.

Obviously, even a clear spheroidal galaxy needs a sufficiently massive disc component to develop a bar, and this component is easily visible in Fig.~\ref{fig:bar_ex}.

\begin{figure*}
    \centering
    \includegraphics[width=0.88\textwidth]{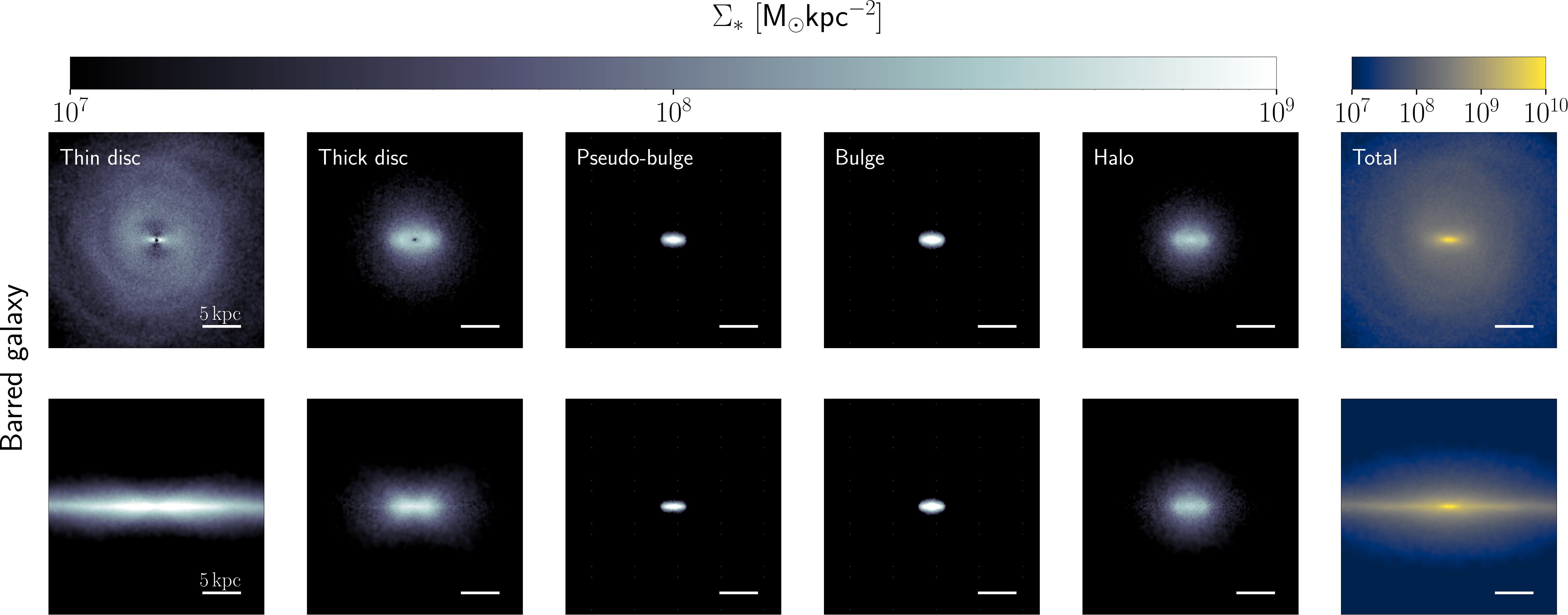}\\
    \caption{Same as Fig.~\ref{fig:density_maps}, for the barred disc galaxy BG, shown in Fig.~\ref{fig:A2prof_ex}.}
    \label{fig:bar_density_maps}
\end{figure*}

As an example, in Fig.~\ref{fig:bar_density_maps}, we decompose the barred disc galaxy BG shown in Fig.~\ref{fig:A2prof_ex}.
Particles within the extent of the detected bar are identified by \code{} as belonging to all the 5 morphological components.
In particular, the bulk of bar particles seems to be included in the bulge and the pseudo-bulge, given the higher radial velocities with respect to those orbiting in the disc.
Moreover, we carved out a cuboid with dimensions $R_{\Phi}\times~1$~kpc~$\times~1$kpc, aligned with the bar, and found this structure to contain at most $8$~percent of the total stellar mass.

Differently from \citet{Du_et_al_2019}, we do not find the bar to significantly affect our morphological distinction between disc-dominated and spheroid-dominated galaxies, mainly because they are correctly identified as mildly rotating components, as expected in the current picture of secular bulge formation processes \citep[e.g.][]{Kormendy_Kennicutt_2004}.

\subsubsection{Redshift dependence}

\begin{figure*}
    \centering
    \includegraphics[width=0.83\textwidth]{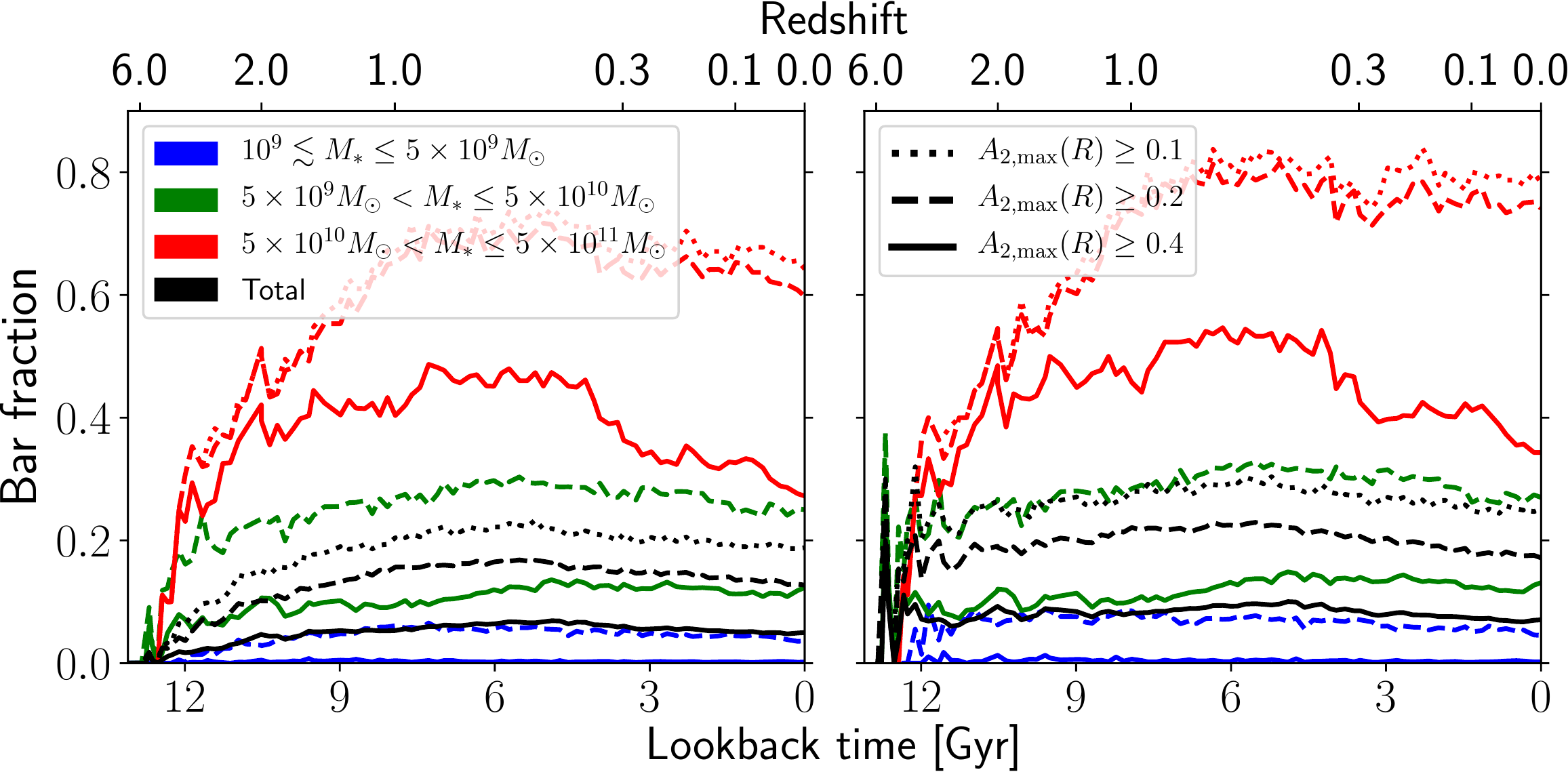}\\
    \caption{Redshift evolution of the bar fraction for all the analysed galaxies ($k_{\rm rot}\geq0.4$; {\it left panel}) and for disc galaxies only ({\it right panel}), according to the definition $M_{\rm thin}+M_{\rm thick}+M_{\rm pseudo-bulge}>0.5M_{*}$.
    The same mass bins of Fig.~\ref{fig:component_fraction} are adopted here, with the same colour code.
    Solid lines correspond to strong bars only ($A_{2, {\rm max}}(R)\geq0.4$), whereas dashed lines include all (proper) bars ($A_{2, {\rm max}}(R)\geq0.2$).
    In the highest mass bin (red lines) and in the total (black lines), we also include the weakest proto-bars found via \code{} ($A_{2, {\rm max}}(R)\geq0.1$).}
    \label{fig:bar_fraction}
\end{figure*}

Fig.~\ref{fig:bar_fraction} shows the result of our bar identification algorithm in the same mass bins used so far and in two different galaxy samples; i.e. the entire population (left panel), and disc galaxies only (fiducial definition; right panel).
We show both strong bars with $A_{2, {\rm max}}(R)\geq0.4$ (in agreement with \citealt{Rosas-Guevara_et_al_2022}; solid lines) and the broader sample of weak-intermediate bars with $A_{2, {\rm max}}(R) \geq 0.2$ (dashed lines), where $A_{2, {\rm max}}(R) = 0.2$ is the minimum threshold usually adopted in literature to identify a two-fold non-axi-symmetry as a bar.
Our code is able to recognize even weaker structures with $0.1\leq A_{2, {\rm max}}(R) < 0.2$ (dotted lines), disentangling them from other density fluctuations, such as growing spiral arms, stellar clusters, or satellite fly-bys, through a simultaneous analysis of both the phase $\phi(R)$ and the strength profile.
For the sake of clarity we show the cumulative sample of bars with $A_{2}(R)\geq0.1$ only for the most massive galaxies and for the total sample. 
All barred galaxies represented in Fig.~\ref{fig:bar_fraction} have a minimum length equal to the equivalent-Plummer of the softening length $R_{\rm min}(z)=2.8h(z)$, at redshift $z$.
Seemingly, both panels show a similar evolution of the bar fraction in all the mass bins, although disc galaxies predictably reach larger values.
Bars ($A_{2, {\rm max}}(R)\geq0.2$) in dispersion-dominated galaxies are unsurprisingly the minority of the total bar sample, reaching up to $\approx 20$ percent in most massive galaxies at $z=0$ (after exceeding $40$~percent at higher redshift), whereas they remain below 7~percent when considering the entire mass spectrum.

The higher the mass of the galaxy, the higher the probability for a galaxy to host a bar, hinting to a major role played by the galaxy self-gravity in triggering the bar instability \citep[in agreement with, e.g.,][]{Peschken_Lokas_2019, Zana_et_al_2019, Rosas-Guevara_et_al_2020}.
We note that at $z\gtrsim6$, galaxies are in general very irregular and too turbulent to maintain a stable long-lived non-axisymmetry.
For this reason, a sudden fluctuation in the stellar surface density, with an almost regular phase, could be interpreted as a bar structure even if it had a different dynamical origin.
However, we stress that our code is not easily deceived by these transient thickenings of the stellar component.
After an initial increase of the bar fraction, the epoch around $z=2$ seems to mark a break in the growth for almost all the subsets of objects we analyse.
This is compatible with bars as naturally arising in unstable galaxies but inhibited and slowed in their growth by tidal interactions with the environment, which are frequent and intense around $z=2$.
At lower redshifts, the bar fraction does not change appreciably, with the exception of a visible decrease near $z=0$, for the most massive systems, where other processes such as bar suicide and buckling would start to disassemble the most mature and evolved sub-structures.
We highlight that, only in the mass bin $5 \times 10^{10} < M_{*}/\msun \leq 5\times10^{11}$, the bar fraction shows a quite different trend when strong bars are studied separately.
As a matter of fact, strong bars in massive galaxies show a clear drop toward low redshifts ($z\leq0.3$).

In summary, at $z=0$, we find 27~(60)~percent of massive galaxies to have a strong (at least a weak-intermediate) bar, whereas this percentage drops to 0.2~(4)~percent for the lowest mass galaxies.
Intermediate-mass galaxies show intermediate values, 12~percent for strong bar and 25~percent for weaker structures.
The total population has instead a cumulative fraction of $5$~percent of strongly barred galaxies and $13$~percent of barred galaxies with $A_{2, \rm max}(R)\geq 0.2$.
If we include even the ``proto-bars'', the ratio reaches almost one fifth of the total population.
Among disc galaxies only -- according to our fiducial definition -- the fractions of strong bars are $34$, $13$, and $0.2$ percent for high-mass galaxies, intermediate mass galaxies, low-mass galaxies, respectively, whereas the total population has an average of $7$~percent.
When all the barred galaxies, with $A_{2, \rm max}(R)$ above or equal to 0.2 are considered, these numbers raise to 74, 27, 5, and 17~percent, and to 78, 38, 11, and 25~percent if bars with $A_{2, {\rm max}}(R) \geq 0.1$ are considered in the calculations.
Galaxies with $5 \times 10^{11} < M_{*}/\msun < 6 \times 10^{12}$ (here included only in the total sample) have a bar fraction lower than the highest mass bin, but larger than the lowest-mass galaxies.
The number of disc galaxies amongst these heavy systems is too low to produce a meaningful estimate.

Our results for the most massive galaxies agree very well with numerous observational studies where little or no evolution of the bar fraction is detected in the redshift range $2\lesssim z \lesssim 0.5$ \citep[][]{Sheth_et_al_2008, Kraljic_et_al_2012, Melvin_et_al_2014, Simmons_et_al_2014}.
However, whereas in our data a slightly decreasing trend -- obvious in strong bars and barely noticeable in the whole sample -- is present for $z\lesssim0.3$, \citet{Sheth_et_al_2008} and \citet{Kraljic_et_al_2012} observe a clear growth until $z=0$.
This discrepancy is possibly due to a missing barred population in observations at $z>0.3$, likely caused by an observational bias.
For instance, our intermediate-mass bin is already too under-massive to be compared with a statistically significant observational counterpart, given the sensitivity and resolution thresholds of current telescopes.
%
%
We refer to the discussion in \citet{Rosas-Guevara_et_al_2022}, where theoretical predictions are reconciled to observed trends when observational biases are taken into account.

\subsubsection{Mass dependence}

\begin{figure*}
    \centering
    \includegraphics[width=0.83\textwidth]{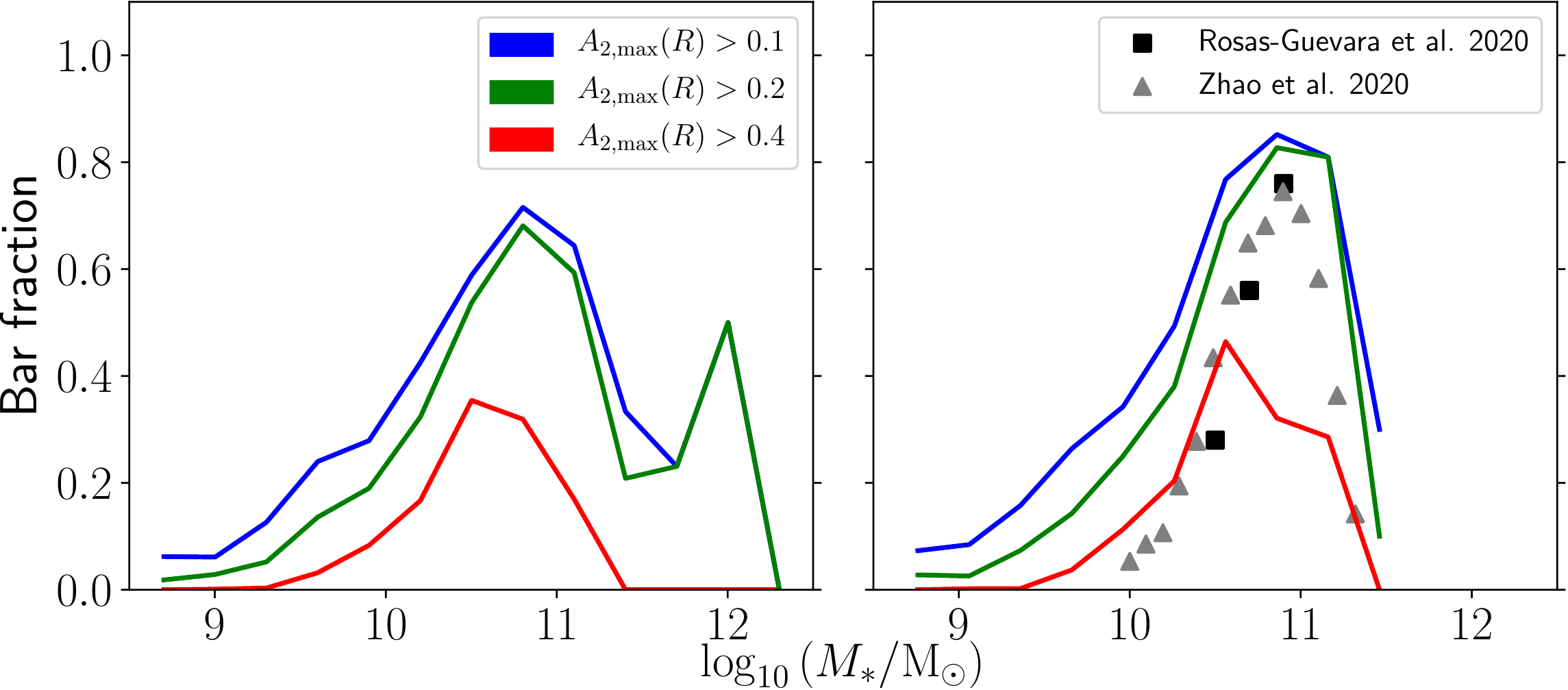}
    \caption{$z=0$, bar fraction as a function of the stellar mass.
    As in Fig.~\ref{fig:bar_fraction}, we compute the bar fraction over total galaxy sample in the {\it left panel} and, only among disc galaxies (i.e., if $M_{\rm thin}+M_{\rm thick}+M_{\rm pseudo-bulge}>0.5M_{*}$), in the {\it right panel}.
    Red, green, and blue lines refer to strong bars only ($A_{2, {\rm max}}\geq0.4$), strong and intermediate bars ($A_{2, {\rm max}}\geq0.2$), and all the bars found ($A_{2, {\rm max}}\geq0.1$), respectively.
    Black squares \citep[][]{Rosas-Guevara_et_al_2020} and grey triangles \citep{Zhao_et_al_2020b} in the right panel show the total bar fraction among TNG100 disc galaxies.}
    \label{fig:bar_fraction_mass}
\end{figure*}

Fig.~\ref{fig:bar_fraction_mass} shows the $z=0$ bar fractions with respect to the total galactic stellar mass, both for the whole sample (left) and for disc galaxies only (right).
We notice that the bar fraction, regardless of the bar strength, increases with mass, and reaches a peak at $M_{*} \lesssim 10^{11}\msun$ in both samples.
Strong bars are always present in the mass range $3\times10^{9}\lesssim M_{*}/\msun \lesssim 3\times10^{11}$, and populate almost 50~percent of disc galaxies in the mass bin around $4\times10^{10}~\msun$.
Weakest bars ($A_{2. {\rm max}} < 0.2$) are mostly found in small galaxies ($M_{*}\lesssim10^{10}~\msun$). 
This can be explained by the fact that, while bar seeds can very easily emerge when the galactic potential is bar-unstable or when it is perturbed -- even only marginally \citep[see, e.g.,][]{Zana_et_al_2018b} -- a fully evolved bar needs a sufficiently massive stellar disc to grow in strength and size via its own self-gravity.
\noindent It is interesting to compare our results with the $z=0$ bar fraction of massive disc galaxies in TNG100. In the right panel of Fig.~\ref{fig:bar_fraction_mass}, black squares show the bar fraction ($A_{2, {\rm max}}\geq0.2$) computed by \citet{Rosas-Guevara_et_al_2020}, via a similar procedure to ours, i.e. a fourier decomposition of the stellar surface density.
On the other hand, grey triangles report the data by \citet{Zhao_et_al_2020b}, where an ellipse-fitting analysis of the surface density map has been adopted.
We note that both the TNG100 trends are in good agreement with our green line, though slightly shifted toward higher masses. Indeed, at fixed mass, a better resolution favours the formation of sub-structures.

Various observational works find an increasing trend of the bar fraction with respect to the stellar mass up to $M_{*}\sim10^{11}~\msun$ \citep[][]{Masters_et_al_2012, Melvin_et_al_2014, Gavazzi_et_al_2015}.
Consistently, other studies indicate the presence of a peak in the distribution for $M_{*}\lesssim10^{11}~\msun$ and possibly the beginning of a decreasing trend \citep[][]{Diaz-Garcia_2016, Consolandi_2016, Cervantes-Sodi_2017}.
All these studies report very similar values ($\sim40-50$~percent) for the bar fraction at $M_{*}\sim10^{11}~\msun$, but no consensus appears at lower masses ($10-50$~percent at $M_{*}\sim10^{10}~\msun$).
Interestingly, \citet{Erwin_2018} suggest an even higher bar faction in their sample ($\sim70$~percent at $10^{10}~\msun$) with respect to our data and see a clear decline at higher masses.
Similarly to our results, the authors claim that the inclusion of S0 galaxies would reduce the steepness of the slope after $M_{*}\sim10^{11}~\msun$, which we can observe in the left panel of Fig.~\ref{fig:bar_fraction_mass}, where we include ``non-disc galaxies''.

\subsubsection{Evolution of bar properties}

\begin{figure}
    \centering
    \includegraphics[width=0.45\textwidth]{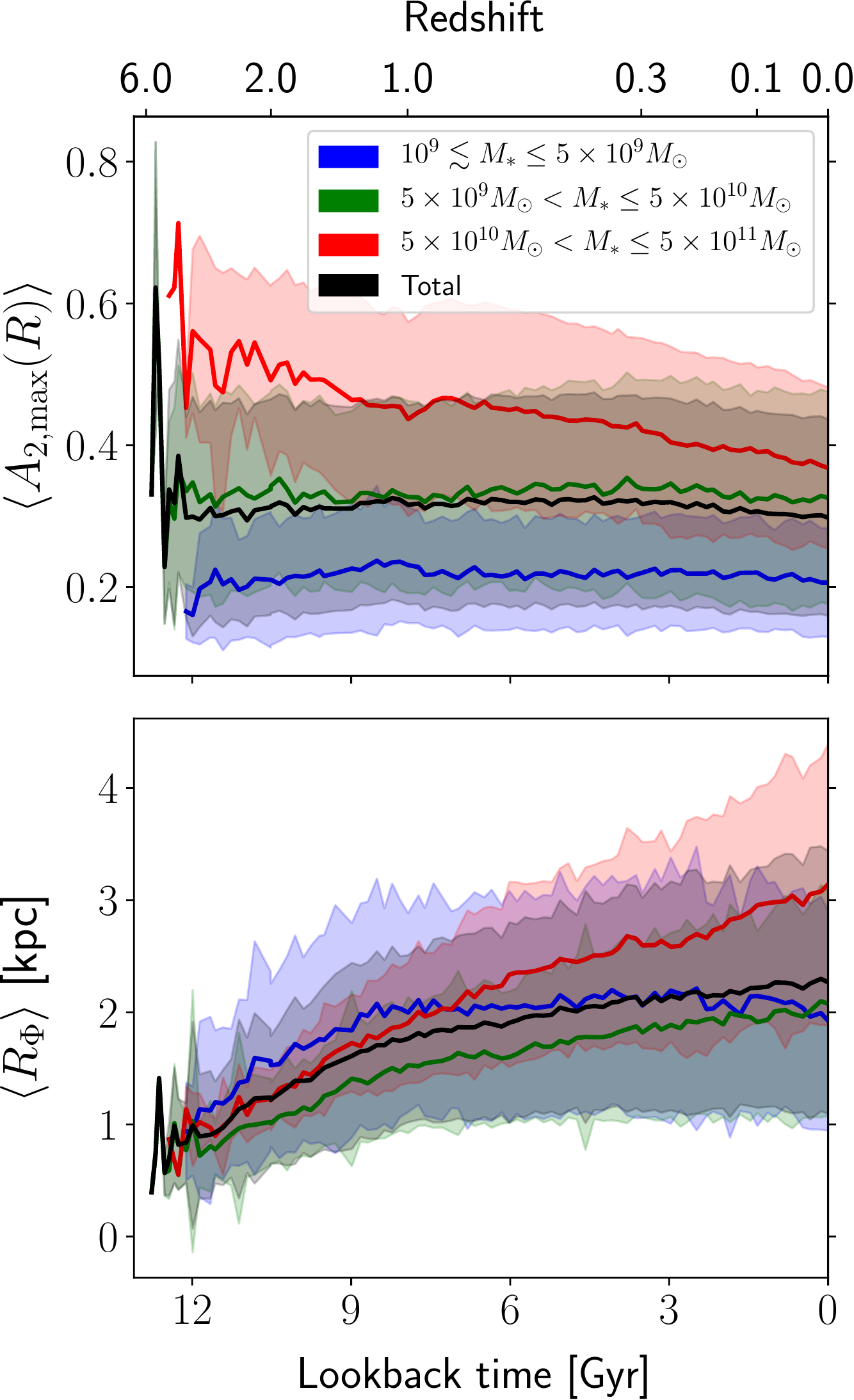}
    \caption{Redshift evolution of the mean bar strength ($\langle A_{2, \rm max}(R) \rangle$; {\it top panel}) and mean bar length ($\langle R_{\Phi} \rangle$; {\it bottom panel}) for the whole galaxy population.
    Shaded areas refer to 1 $\sigma$ of each distribution.}
    \label{fig:mean_bar_prop}
\end{figure}
In Fig.~\ref{fig:mean_bar_prop} we show the redshift evolution\footnote{Similarly to \S~\ref{subsec:redshift_evolution}, we remark that we do not refer here to the proper evolution of bars inside their host galaxies, but, rather, to the changes of the average bar properties at different redshifts.} of the main bar properties computed via \code{}, namely the strength (via the value $A_{2, \rm max}(R)$; top panel) and the extent (via the radius $R_{\Phi}$; bottom panel).
We do not report in the Figure the mean properties of bars hosted in disc galaxies only, since they do not show any appreciable difference with respect to the total population.
%
%
Only the most massive galaxies (red line) show a clear evolution in the strength, whereas lower-mass galaxies maintain an average constant value of $A_{2, \rm max}(R)$ for the entire cosmological history.
This is in agreement with a recent paper by \citet{Kim_et_al_2021} where the authors analyse the bar properties of almost 400 galaxies in the redshift range $0.2<z<0.8$.
Interestingly, it seems that bars struggle to increase their strength above a certain stellar-mass-dependent threshold.

On the contrary, the mean bar length grows almost monotonically in each mass bin, with the only exception of the least massive galaxies (blue line) at low redshift.
Although with a milder dependence with respect to the bar strength, a correlation between the average bar size and the galactic mass is still visible, in agreement with \citet{Kim_et_al_2021}.
The most massive galaxies show the largest variation in their evolution, whereas lower-mass galaxies seem to saturate to a mean value of about $2$~kpc.
Unfortunately, the exact magnitudes of the bar size can be hardly compared with other observational and numerical works, given a strong dependence on the method adopted \citep[see, e.g.,][]{Athanassoula_Misiriotis_2002}.
For example, whereas our mean bar length is in good agreement with \citet[][]{Gadotti_2009}, we find shorter values with respect to \citet{Erwin_2018} and \citet{Kim_et_al_2021}.



\section{Discussion and conclusions}
\label{sec:conclusions}

In this paper, we presented \code{}, a new galaxy kinematic decomposition algorithm, that we applied to the whole galaxy sample in the TNG50 cosmological simulation \citep[][]{Pillepich_et_al_2019, Nelson_et_al_2019a}.
Our method uses a differential classification of the energy and circularity of each particle associated to a selected galaxy. 
Since galaxies extracted from cosmological simulations are generally much ``messier'' than those of isolated simulations, we needed to develop a number of dedicated routines (as outlined in Section~\ref{sec:method}) to systematically and correctly classify particles.
We compared our findings with those obtained by other methods available in the literature, finding a generally good agreement, but also demonstrating the greater flexibility and consistency of our approach.
\code{} successfully decomposes all the galaxies in the sample without producing any outliers, differently from the other methods (zero-mass discs, cumulative mass fraction of the components larger than unity, etc.).

In addition to our five galactic components (bulge, halo, pseudo-bulge thin and thick disc), we also recognised galaxies that show a bar structure. Remarkably this feature is not exclusive of galaxies classified as discs, mostly due to the somewhat arbitrary definition of ``disc galaxy'' that would not include galaxies with a subdominant disc component or that were disc dominated in their recent past\footnote{We stress that a measurable disc component is present any time a bar is observable, even if in a minority of cases a bar is observed in non-disc-dominated galaxies; see e.g. the  galaxy in  the top row of Fig.~\ref{fig:bar_ex}.}.
\code{} is able to identify even the weakest bars thanks to the simultaneous analysis of the strength and phase of the two-fold non-axisymmetries, making the procedure very suitable for studying the onset of the bar formation process.
Unfortunately, without a proper analysis of the orbits, which would be unrealistic in cosmological simulations of this size, it is extremely challenging to assign star particles to a separate and specific bar component.
Hence, stars orbiting in the bar inevitably pollute the standard morphological components. We proved that the whole classification process is not affected by the presence of these perturbations from axi-symmetry.

We take advantage of the versatility of \code{} to analyse all the redshifts of TNG50 and provide the morphology catalogue for both barred and unbarred systems.
Such a catalogue gives the possibility to perform a detailed study on the evolution and formation of these structures across the entire simulation.
We found an initial increasing $D/T$ ratio and later flattening of the number of disc galaxies at all the galactic masses, except for the most massive ones ($5 \times 10^{10} < M_{*}/\msun \leq 5 \times10^{11}$), which we expect to evolve in the most dispersion dominated systems of the cosmological volume.
We show a growing bar fraction ($A_{2, {\rm max}}(R) \geq 0.2$), which eventually saturates below $z\sim2$ in the entire galaxy mass range and appreciably decreases only in the most massive sample, being -- at $z=0$ -- about 0.04, 0.3, and 0.6 for small, intermediate, and massive galaxies, respectively, and 0.05, 0.3, 0.7 in disc galaxies. The average bar strength decreases with redshift only for most massive galaxies, while the average bar length always increases for all galaxies.
In agreement with observations, more massive galaxies are favoured as bar-hosts and, at $z=0$, about the majority of systems with $M_{*}\gtrsim10^{11}\msun$ are barred, with a clear peak around $M_{*}\sim10^{11}~\msun$ in agreement with numerous observational works. Such result hints to a major role of this substructure in the evolutionary process of the most massive galaxies.

We have shown that \code{} is extremely versatile and reliable in determining the kinematic and morphological properties of simulated galaxies, and this makes it the perfect tool to investigate the dynamical evolution of galaxies across cosmic time.


\section*{Acknowledgements}
TZ warmly thank Min Du and collaborators for providing the data of their morphological decomposition and Annalisa Pillepich for the useful comments and suggestions. AL, MB, and MD acknowledge funding from MIUR under the grant PRIN 2017-MB8AEZ.
YRG acknowledges the support of the ``Juan de la Cierva Incorporation'' fellowship (IJC2019-041131-I) and the European Research Council through grant number ERC-StG/716151.
DN acknowledges funding from the Deutsche Forschungsgemeinschaft (DFG) through an Emmy Noether Research Group (grant number NE 2441/1-1).
The primary TNG simulations were carried out with compute time granted by the Gauss Centre for Supercomputing (GCS) under Large-Scale Projects GCS-ILLU and GCS-DWAR on the GCS share of the supercomputer Hazel Hen at the High Performance Computing Center Stuttgart (HLRS).
We greatly thank the anonymous referee for the useful comments which improved the quality of this manuscript.

\section*{Data Availability Statement}
\code{} can be found at \href{https://github.com/thanatom/mordor}{https://github.com/thanatom/mordor}.
The whole TNG50 catalogue is publicly available at \href{www.tng-project.org/zana22}{www.tng-project.org/zana22}.
The IllustrisTNG simulations are publicly available and accessible in their entirety at \href{www.tng-project.org/data}{www.tng-project.org/data} \citep{Nelson_et_al_2019a}.
The remaining data underlying this article will be shared on reasonable request to the corresponding author.


\bibliographystyle{mnras}
\bibliography{bibliography}


\appendix

\section{Potential evaluation modes}
\label{sec:potential_details}

\begin{figure}
    \centering
    \includegraphics[width=0.45\textwidth]{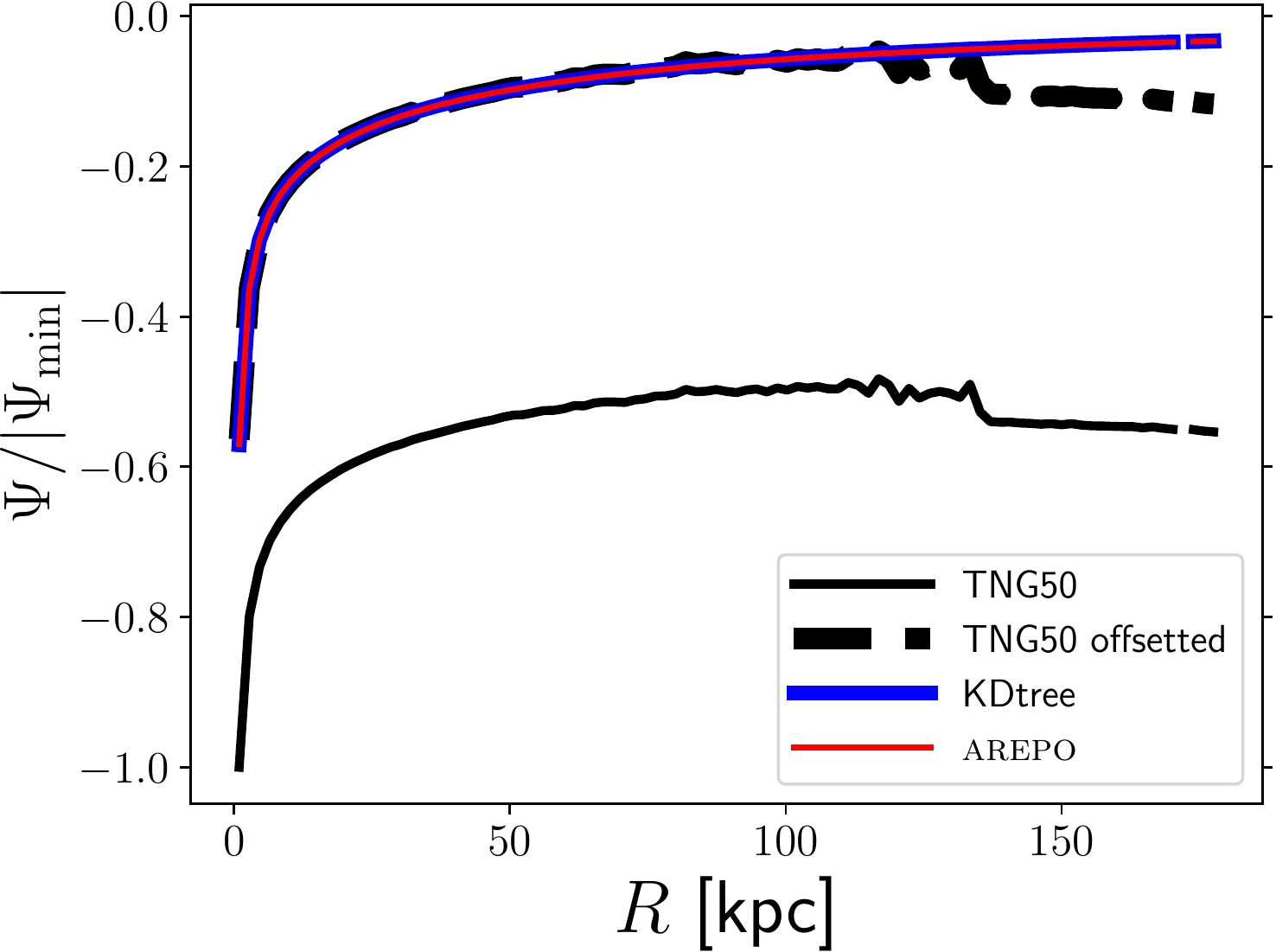}\\
    \caption{Radial profiles of the gravitational potential energy per mass unit of a $6.8\times10^{10}\msun$ galaxy, evaluated through different techniques.
    Black lines show the direct output of the TNG50 simulation before (solid line) and after (dashed line) the rigid offset is applied.
    Blue and red lines show the outputs of the KDtree and \textsc{arepo} code, respectively.}
    \label{fig:pot_prof}
\end{figure}

In order to evaluate the potential for each galaxy particle, our tool provides several alternatives ways, listed below.
Three of them are exemplified in Fig.~\ref{fig:pot_prof}.
\begin{itemize}
    \item \textbf{Simulation mode (cosmological simulation)}: the potential is directly read from the simulation snapshots (if available). Although this approach is the most straightforward (when the potential is a direct output of the simulation; solid black line in Fig.~\ref{fig:pot_prof}) one should keep in mind that the energy of the circular orbit estimated in our code to determine the circularity only accounts for the particles associated with the galaxy to examine, without the global contribution of the cosmological large-scale structure, and this would result in a discrepancy that should be corrected.
    For this reason, when this option is chosen, we apply a rigid offset to the particle potential so that the radial profile from the simulation closely matches the profile estimated by our code in normalisation (dashed black line);
    \item \textbf{Simulation mode (non-cosmological)}: the potential is still directly read from the simulation snapshots, but with the potential corresponding to that of an isolated system. While this is straightforward in isolated galaxy simulations, for cosmological runs the potential must be separately recomputed on each galaxy by the hydrodynamic code (\textsc{arepo} in this case; red line in Fig.~\ref{fig:pot_prof}), accounting for the galaxy particles only. This is the approach we followed for our analysis, because of the large number of galaxies to analyse, and the strongly parallelised nature of the hydrodynamic code, that improved the analysis performance.
    \item \textbf{Particle-Mesh mode}: the potential is computed via the Fast Fourier Transform of the particle density field on a Cartesian uniform grid. Although very easy to use, its accuracy strongly depends on the number of cells employed, and can become quite memory-expensive for very fine grids, typically required to accurately resolve the potential gradient in the galaxy centre;
    \item \textbf{Direct summation mode}: the potential is recomputed via direct summation. This is the most CPU-expensive way available, and has been included for completeness, since it is the most accurate method, although it becomes computationally challenging for objects with more than 10$^{5}$ particles;
    \item \textbf{Tree mode}: the potential is recomputed via a KD tree (blue line in Fig.~\ref{fig:pot_prof}), that is directly coupled with our code. This option is based on a suitably modified version of \textsc{pytreegrav}\footnote{\url{https://github.com/mikegrudic/pytreegrav}.} and gives very accurate results at a reasonable computational cost, especially because of the optimisation performed with \textsc{numba}.
    However, it can nevertheless become quite expensive for galaxies with more than $10^{7}$ particles, given very basic multi-threading parallelisation.
\end{itemize}


\bsp	
\label{lastpage}
\end{document}